\documentstyle[epsf,psfig,aps]{revtex}
\renewcommand{\thefootnote}{\fnsymbol{footnote}}
\begin{document}
\newcommand{\Le}{\left(}
\newcommand{\Ra}{\right)}
\newcommand{\be}{\begin{equation}}
\newcommand{\dlq}{\lq\lq}
\newcommand{\ee}{\end{equation}}
\newcommand{\ben}{\begin{eqnarray}}
\newcommand{\een}{\end{eqnarray}}
\newcommand{\stackeven}[2]{{{}_{\displaystyle{#1}}\atop\displaystyle{#2}}}
\newcommand{\lsim}{\stackeven{<}{\sim}}
\newcommand{\gsim}{\stackeven{>}{\sim}}
\renewcommand{\baselinestretch}{1.0}
\newcommand{\as}{\alpha_s}
\def\eq#1{{Eq.~(\ref{#1})}}
\def\fig#1{{Fig.~\ref{#1}}}
\begin{flushright}
Brown--HET--1362 \\
TAUP--2739--2003\\
\end{flushright}
\vspace*{1cm} 
\setcounter{footnote}{1}

\begin{center}
{\Large\bf High energy amplitude as an admixture of `soft' and `hard' Pomerons}
\\[1cm]
Sergey Bondarenko $^{1}$\,,Eugene Levin $^{1}$ \,and\, C-I Tan $^{2}$  \\ ~~ \\
{\it $^1$ HEP Department, School of Physics and Astronomy } \\
{\it Tel Aviv University, Tel Aviv 69978, Israel } \\ ~~ \\
{\it $^2$ Physics Department, Brown University, } \\ 
{\it  Providence RI 02912, USA } \\ ~~ \\ ~~ \\
\end{center}

\begin{abstract} 
In this paper an attempt is made to find an interface of the perturbative BFKL 
Pomeron with the
non-perturbative Pomeron originating from non-perturbative QCD phenomena such
as QCD instantons and/or scale anomaly. The main idea is that the non-perturbative
Pomeron involves a large scale ($M_0 \,\approx\,2\,GeV $), which is larger than the scale
from which perturbative QCD is applicable. One key  result is that even for
processes involving a large hard scale ( such as DIS ) the low $x$ behavior is
determined by an effective Pomeron with an intercept having an essential non-perturbative
QCD contribution. 
\end{abstract}

\renewcommand{\thefootnote}{\arabic{footnote}}
\setcounter{footnote}{0}

\section{Introduction}
In this paper we present an attempt to solve one of the most challenging 
problems of QCD: the interface  between long distance  (non-perturbative, 
soft) interaction and the short distance one (perturbative, hard) at 
high energy.  The short distance interaction at high energy  is controlled 
by perturbative QCD  and can be described by the BFKL equation 
\cite{BFKL}. We call  the hard Pomeron a solution to this equation, 
in spite of the fact that this  solution is not a Regge pole.
The region of long distances we describe by introducing the 
phenomenological soft Pomeron - a Regge pole with the intercept close to 1
\cite{DL}.  Recently, progress has been achieved in a 
theoretical understanding of the structure of the soft Pomeron 
\cite{KL,KKL,S,KA,JP,BK,CT}, which provides hope for developing a first 
approach to  understand the interface of these two Pomerons: the soft Pomeron and 
the BFKL one.

\begin{figure}
\begin{center}
\epsfxsize=12cm
\leavevmode
\hbox{ \epsffile{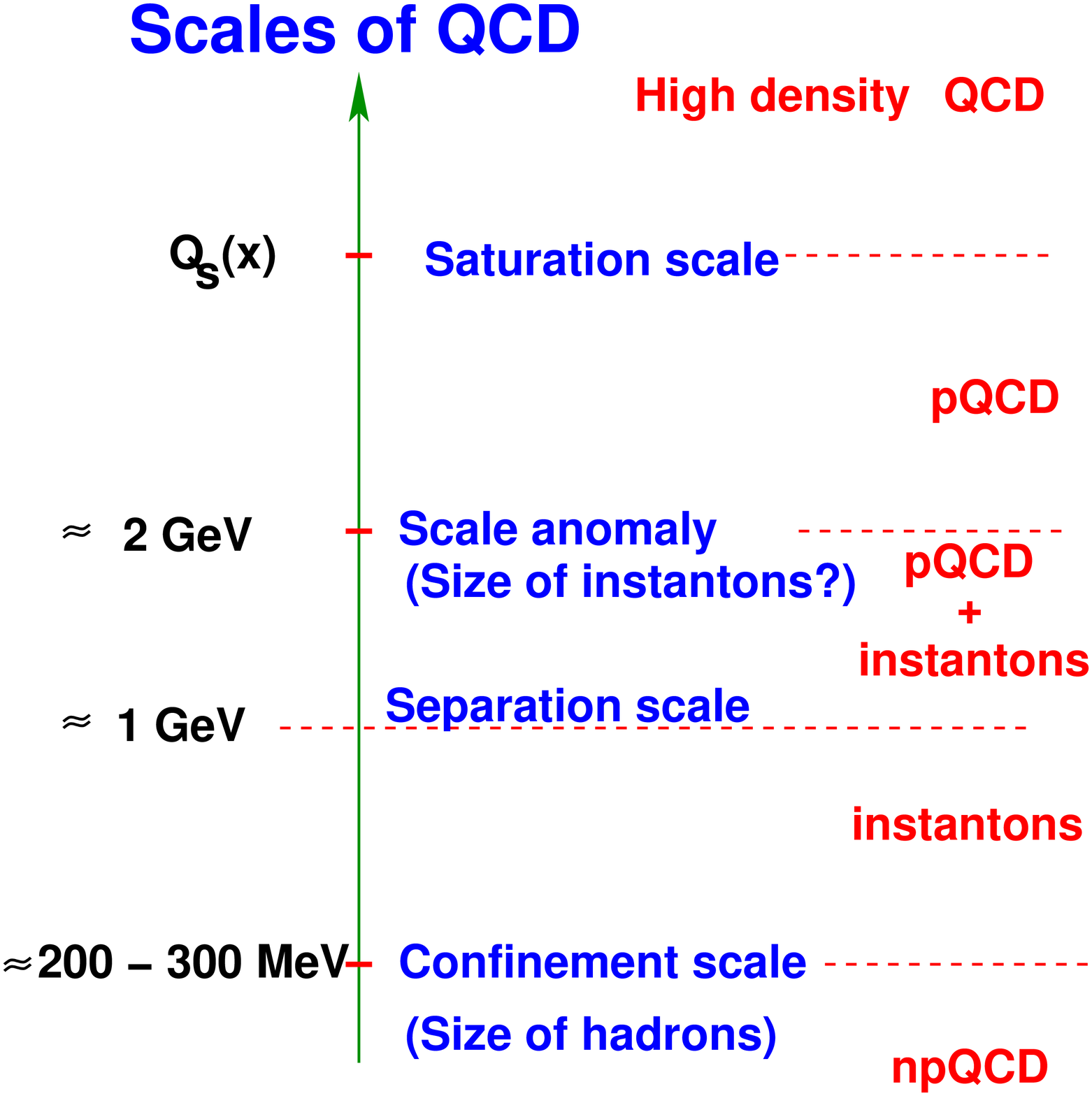}}
\end{center}
\caption{Scales of QCD accordingly to  Ref. \protect\cite{KL}.  }
\label{scale}
\end{figure}

This approach is based on a new hierarchy of the scales in QCD suggested in 
Ref. \cite{KL} and depicted in Fig.~\ref{scale}.

The key idea underlying this picture is the large scale for the soft Pomeron 
($ q_S=M_0 \, \approx \,2\,GeV$) which is closely related to the violation 
of the scale anomaly of QCD. The value of $M_0$ is so large that  
the QCD coupling constant $\alpha_S(M_0)\,\ll\,1$, and it can be considered 
to be a small parameter. From Fig.~\ref{scale} we see that  such  
sequence of scales makes possible a theoretical approach to the problem 
of interface of soft and hard Pomerons. For all virtualities 
(transverse momenta) larger than the separation scale, $q\,>\,q_{H}\,\approx\,1\,GeV$,
 where $\alpha_S(q^2) \,<\,1$, we can apply the perturbative QCD approach.

We formulate the problem of the interface between soft and hard
Pomeron as a problem of a joint description of the perturbative BFKL Pomeron with the 
soft
Pomeron based on the non-perturbative contributions with sufficiently high
scale (see Fig.~\ref{scale}). Many  of the  non-perturbative
contributions originate from very long distances of the order of the
confinement scale (see Fig.~\ref{scale}) but we assume that all of them are
irrelevant to the structure of the soft Pomeron \cite{KL}.

We also assume that the energies are so high that the saturation scale 
\cite{GLR,SS} is much larger than the scale of the anomaly scale, 
(or the scale of the soft Pomeron),
$q_S\, \equiv \,M_0\, \approx \,2\,GeV$ and the separation scale $q_H$ which
characterizes that value of the virtuality (transverse momentum) 
from which we can start using perturbative QCD. Our notation is transparent since
$q_S$ gives the scale of the soft Pomeron, while for $q \,> \,q_H$ we can use the
BFKL Pomeron to describe the high energy scattering in pQCD.

In the next section we describe the general formalism of our approach
which leads to a new Regge pole as a result of an admixture  of  soft and hard
Pomerons with the trajectory $ \alpha_{S-H}(t)
\,=\,1\,+\,\Delta_{S-H}\,+\,\alpha'_{S-H}\,t$. Section 3 is devoted entirely to
the properties of the Green's function of the BFKL Pomeron with the running
QCD coupling  constant. As it has  been shown in many papers
\cite{GLR,CK,EL,KM,ABB,EL1,CCS,CTM} the QCD running  coupling constant changes the
character of the singularity in the BFKL Pomeron which becomes an essential
singularity at $\omega \,\rightarrow\,0$ instead of a pole at $\omega =
\omega_L$.  In this section we present a detailed analysis of the Green's
function for the BFKL equation which is similar to that of
 paper \cite{CTM} which appeared after we had completed our calculations.

In section 4 we calculate the trajectory of the soft Pomeron which leads to
the phenomenological Pomeron ($\alpha_{phenom. P}\,\,=\,\,1 \,+\,0.08
\,+\,0.25\,GeV^{-2} t$ \cite{DL}) and which appears as a result of the admixture
of the BFKL and soft Pomerons in our approach.  The deep inelastic scattering
in the region of rather low virtualities of photon is discussed in section 5
where the values of virtualities for the hard processes are specified using
our approach for admixture of ``soft'' and ``hard'' Pomerons.

\section{General Approach}

In this section we outline the general approach for dealing with the interface of soft
and hard Pomerons. To do this we need to  specify our description of the soft
Pomeron based on general ideas of Ref. \cite{KL}. We next review the key feature of BFKL Pomeron in the LLA, elaborate
its Greens' function in the Airy model, and finally discuss unifying soft and hard Pomerons.

\subsection{Soft Pomeron }

\begin{figure}
\begin{center}
\epsfxsize=12cm
\leavevmode
\hbox{ \epsffile{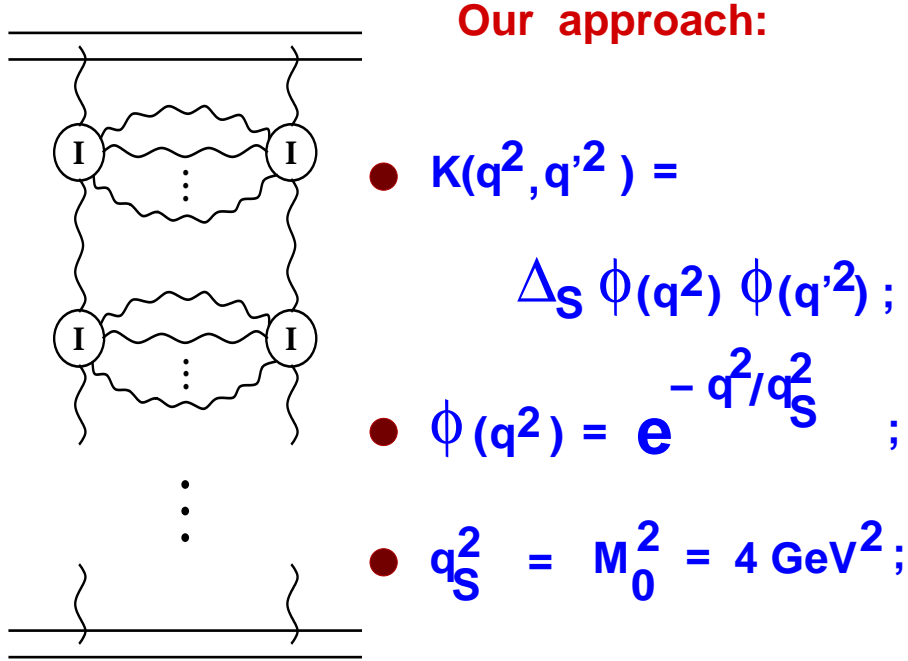}}
\end{center}
\caption{The main properties of the soft Pomeron originating from the QCD
instanton approach \protect\cite{KKL}  }
\label{SPom}
\end{figure}

As a  working model for the soft Pomeron we shall adopt the QCD instanton approach
\cite{KKL}. This is in fact not necessary since we only need to use  general 
properties of the soft Pomeron exchange depicted in Fig.~\ref{SPom}.

As  shown in Ref.\cite{KKL}, the structure of the soft Pomeron in the
 instanton approach formally reduces to summing the ladder diagrams of
Fig.~\ref{SPom}. These ladder diagrams describe the $t$-channel gluon that
propagates through the space inducing a chain of instanton transitions 
between different vacua and producing a number of gluons in each transition.
The sum of these diagrams leads to power-like (Regge-type) asymptotic amplitude, namely
the scattering amplitude $A(s,t)$ is asymptotically equal to
\be \label{SP0}
A(s,t) \,\,=\,\,g(t) \,s^{\alpha_{SP} (t)}\,\,\,\,,
\ee
with $\alpha_{SP} (t)\,=\,1\,+\,\Delta_S\,+\,\alpha'_S\,t$.
It is well known (see Ref. \cite{RT}) that both $\Delta_S$ and $\alpha'_S$
can be found as solutions of the Bethe-Salpeter type equation resulting from these 
ladder graphs. In this paper, we shall concentrate on the forward 
scattering limit, $t=0$, 
where the kernel $K(q^2,q'^2)$ is a function of transverse momenta,  $q$ and $q'$, 
along adjacent rungs of the ladder in Fig.~\ref{SPom}. We assume 
that this kernel has the general
factorized properties, namely $ K(q^2, q'^2)\, =
\,\Delta_S\,\phi(q^2)\,\phi(q'^2)$ which follow  from the instanton approach
of Ref. \cite{KKL} but this could be more general.

\subsection{BFKL Pomeron}

\begin{figure}
\begin{center}
\epsfxsize=10cm
\leavevmode 
\hbox{ \epsffile{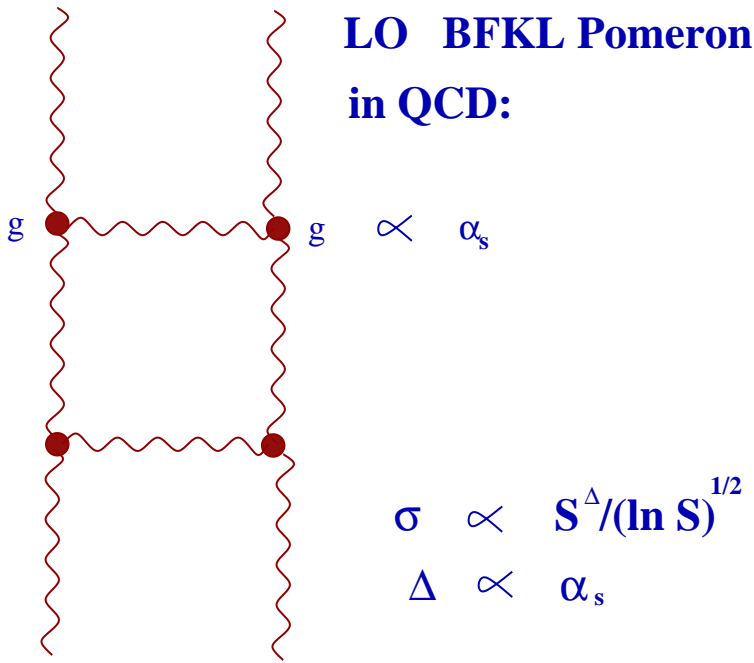}}
\end{center}
\caption{The BFKL Pomeron in LLA.}
\label{BFKL}
\end{figure}

The BFKL Pomeron gives the high energy asymptotic behavior  for the scattering
amplitude in the so-called leading log(1/x) approximation (LLA) of  perturbative
QCD \cite{BFKL}. In the LLA, for each set of Feynman diagrams of the order of $\as^n$, 
one keeps only leading terms proportional to $ (\as
\ln(1/x) )^n$,  and sum over these terms. In LLA this sum formally 
reduces to a summation of ``gluon ladder'' diagrams,  (see Fig.~\ref{BFKL}).  

However, 
due to the  specific property of the vertex\cite{BFKL}, the sum of these ``ladder'' diagrams 
leads to an amplitude $ A(s,t)
\,\propto\, s^{\Delta_H}/\sqrt{\ln(s)}$, with $ \Delta_H \,\propto\,\as$. 
The presence of the additional $1/\sqrt{\ln(s)}$ factor reflects the fact that the BFKL Pomeron
 is actually not a Regge pole.

\begin{figure}
\begin{center}
\epsfxsize=12cm
\leavevmode 
\hbox{ \epsffile{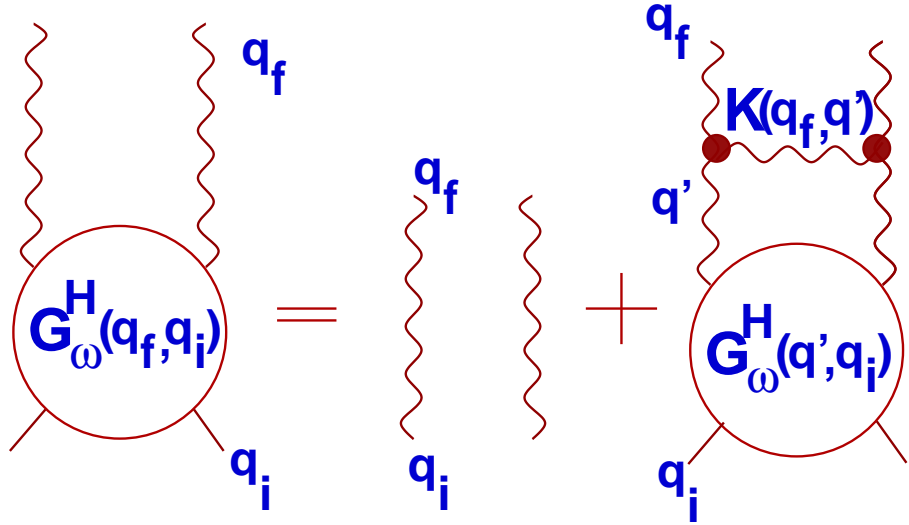}}
\end{center}
\caption{The BFKL equation for the Green function}
\label{GEBFKL}
 \end{figure}

For further discussion, it is more convenient to work in the complex angular momentum plane, 
$l\equiv 1+\omega$,
and to restrict ourselves to the forward limit $t=0$.  In particular, 
 we need to find the Green's function\footnote{We shall also denote this Green's function
as $G^H_{\omega}$ to emphasize that it is responsible
for the ``hard'' perturbative QCD contribution.} $G_{\omega}(\vec{q}_f,\vec{q}_i )$ for the BFKL
equation which satisfies the following equation (see
Refs. \cite{ABB,EL1,CTM} for details and Fig.~\ref{GEBFKL} for graphical form
of the equation)
\be 
\label{GBFKL}
\omega \,G_{\omega}(\vec{q}_f,\vec{q}_i\,\,)=\,\,\delta^{(2)}( \vec{q}_f -
 \vec{q}_i)\,\,+\,\,\int\,d^2
q'\,K(\vec{q}_f,\vec{q}')\,G_{\omega}(\vec{q}',\vec{q}_i)\,\,.
\ee
The kernel  $K(\vec{q}_f,\vec{q}')$ in \eq{GBFKL} can be written, after integration over the azimuthal angle, in the form
\be 
\label{KBFKL}
K(\vec{q}_f,\vec{q}')\,\,=\,\,\as(q^2_f)\,\tilde{K}\left(\frac{q^2_f}{q'^2}\right)\;.
\ee
It can be shown that functions 
\be 
\label{EF}
\phi_f(q^2)\,\,
\,=\,\,\,
\left(\,\frac{q^2}{\sqrt{\as(q^2)}}\,\right)^{f}\;,
\ee
 satisfy a generalized eigenvalue equation
\be 
\label{EV}
\as(q^2)\,
\int\,d q^{'2} \,\tilde{K}\left( \frac{q^2}{q'^2}
\right)\,\,\phi_f(q'^2)\,\,\,=\,\,\,\frac{r_H}{r}
\omega(f)\,\phi_f(q^2)\;,
\ee
with  $\omega(f)$ as eigenvalues. Here, $r \,=\,\ln(q^2/\Lambda^2)
\,\,-\,\,\frac{1}{2}\ln \as(q^2)$, 
 $r_H = \ln(q^2_H/\Lambda^2)$, and $\omega(f)$ in leading order of pQCD is
equal to 
\be 
\label{OMEGA}
\omega(f)\,\,\,=\,\,\,\frac{
\as(r_H) \,N_c}{\pi}\,\{ \,
2\,\psi(1) \,-\,\psi(\frac{1}{2} \,-\,f)\,-\,\psi(\frac{1}{2} \,+\,f)\,\}\;.
\ee
In this paper we will mostly use the expansion of $\omega(f)$ at small values of
$f$, namely,
\be \label{OMEGAD}
\omega(f)\,\,\,=\,\,\,\omega_L
\,\left(\,1\,\,+\,\,D\,\,f^2\,\,+\,\,O(f^4)\,\right)\,\,.
\ee
It should also be stressed that in \eq{OMEGA} and \eq{OMEGAD} the normalization
point for the QCD running coupling constant $\as$ was chosen as the
separation scale, namely $\as(q^2_H)$.

The  solution to the inhomogeneous \eq{GBFKL} can be found by  expanding
$G_{\omega} (r_f,r_i)$ in terms of the complete set of eigenfunctions, \eq{EF},
\ben
&&G_{\omega}(\vec q,\vec q_i) \equiv   G_{\omega}(r,r_i)=
\,\,\int^{a \,+\,i\infty}_{a - i\infty}\,\frac{d f}{2 \pi
i}\,\,\chi( \omega,f,r_i)\,\,
\phi_f(q^2) \,=\,\int^{a
\,+\,i\infty}_{a -i\infty}\,
\frac{d f}{2 \pi i}\,\,\chi( \omega, f,r_i)\,\,e^{r f} \,\,,\label{ME1}\\
&&G(Y;r_f,r_i) = \,\int \,\frac{d\,\omega}{2\,\pi\,i}
\,e^{\omega\,Y}\,G_{\omega}(r_f,r_i)\,\,,  \label{ME2}
\een
where $ Y = \ln(s/q_f\,q_i)$.

Using \eq{ME1} one obtains the following equation for $\chi(\omega,f,r_i)$
\cite{EL1}
\be 
\label{MEBFKL}
-\,\omega \frac{d \chi (\omega, f,r_i)}{d \,f}\,\,=\,\,r_H\,\omega (f)\, \,\chi (\omega,f,r_i)\,\,
+\,\,r_i \,e^{-f\, r_i}\,\,.
\ee
\eq{MEBFKL} is a first order differential equation and  can be solved analytically. 
 The solution to  \eq{MEBFKL} can be written as  the sum of two terms,\cite{EL1}:
\be \label{SOLME}
\chi_{1}(\omega, f,r_i)\,\,=\,\,\frac{r_i}{2\,\omega}\,\int_f \,\,d\,f' \,e^{ -
f'\,r_i\,\,+\,\,\frac{r_H}{\omega}\,\int^{f'}_f
\,\,\omega(f'')\,d\,f''}\,\,,
\ee
and 
\be \label{SOLME1}
\chi_{2}(\omega, f,r_i)\,\,=\,-\,\frac{r_i}{2\,\omega}\,\int^f \,\,d\,f' \,e^{ -
f'\,r_i\,\,+\,\,\frac{r_H}{\omega}\,\int^{f'}_f
\,\,\omega(f'')\,d\,f''}\,\,,
\ee
where the integration path in \eq{SOLME} is a ray $C_1$, from f to infinity, lying
 between the polar angles  $\pi/3$ and $\pi/2$ and the corresponding path for
\eq{SOLME1} is another ray $C_2$ from f to minus infinity,
lying between the polar angles  $-\pi/3$ and $\,-\pi/2$.
For 
$G_{\omega}(r_f,r_i)$, we obtain
\be \label{G1}
G_{\omega}(r_f,r_i)\,\,=\,\,\int^{\tilde{f} + i
 \infty}_{\tilde{f} - i \infty}\,\frac{d f}{2\,\pi\,i}\,\,e^{\,f\,r_f \,\,}\,
\Le\,
\,\chi_{1}(\omega, f,r_i)\,+\,\chi_{2}(\omega, f,r_i)\,\Ra\,\,.
\ee
It is worthwhile mentioning that the difference between 
\eq{SOLME} and \eq{SOLME1} is a solution to the homogeneous BFKL
equation with the running QCD coupling constant, (\eq{MEBFKL} without the
inhomogeneous term). The fact that 
$\chi(\omega, f,r_i)=\chi_{1}(\omega, f,r_i)\,+\,\chi_{2}(\omega, f,r_i)   $  
corresponds to having chosen the  
integration constant to  \eq{MEBFKL}  in order to ensure  that
$G_{\omega}(r_f,r_i)$ is real.

\subsection{The BFKL Green function in the Airy model}

  To clarify the main properties of the Green's function we consider the
expansion of \eq{OMEGAD} which is justified at small values of $f$ and
$f'$. We call this approach the Airy model, and  it was first done in
Ref.\cite{CTM}, our functions will be very
close to the Airy functions. Using new variables $f^+ = f'\,+\,f$ and
$f^-\,=\,f'\, - \, f$, we can reduce \eq{G1} to the form
\ben
G_{\omega}(r_f,r_i)\,\,=
g(\omega,r_f, r_i) &=& \frac{r_i}{4\,\omega}\,\int^{\tilde{f} + i\infty}_{\tilde{f} - i
\infty}\,\frac{d f^+}{2\,\pi\,i}\,\Le\,\int_{C_1}\,d\,f^-
\,\,e^{\Psi(\omega; f^+, f^-; r_f, r_i)}\,
+\int_{C_2}\,d\,f^-\,\,e^{\Psi(\omega; f^+, f^-; r_f, r_i)}\,\Ra\,\label{AM1}\\
\Psi(\omega; f^+, f^-; r_f, r_i) &=& 
-\,\frac{r_f + r_i}{2}\,f^{-} \,\,+\,\,\frac{ r_f -r_i}{2}\,f^{+}
\,\,+\,\,\frac{\omega_L \,r_H}{\omega}\, f^-\,\,\{\,1\,+\,
\frac{D}{3}\,[\,\frac{3}{4}\,( f^{+})^2\,\,+\,\,\frac{1}{4}\,( f^{-}
)^2\,]\,\}\,\,;\label{AM2}
\een
Integration paths $C_1$ and $C_2$
in \eq{AM1} are  defined in the same way 
as in previous section, with the initial point
shifted to $f^-=0$.

Since the $f^+$-integral is Gaussian, we have
\be\label{ORIG}
g(\omega,r_f, r_i)\,\,=\,
\,\sqrt{\frac{r_i^2}{16\,\pi\,\omega\,\omega_L\,r_H\,D}}\,
\Le\,\int_{C_1}\,+\,\int_{C_2}\,\Ra\,
\frac{d\,f^{-}}{\sqrt{f^{-}}}\,\,
e^{-f^{-}\,\frac{r_{f}+r_{i}}{2}\,\frac{\omega\,-\,\tilde{\omega}}{\omega}\,+
\,(f^{-})^{3}\,\frac{D\,\omega_{L}\,r_{H}}{12\,\omega}\,-\,
\frac{1}{f^{-}}(\frac{\delta\,r}{2})^{2}\,\frac{\omega}{D\,\omega_{L}\,r_{H}}\,}\,\,,
\ee
with the $\tilde{\omega}\,=\,\frac{r_H\,\omega_L}{\frac{r_{f}\,+\,r_{i}}{2}}$
and $\,\delta r = r_f \,-\,r_i\,$.
We can further change the variable to: 
$\nu\, = \,\Le\,\frac{4\,\omega}{D\,\omega_{L}\,r_{H}}\,\Ra^{-1/3}\, f^-$ ,
and obtain for our Green's function: 
\ben
g(\omega,r_f, r_i) &=& 
\,\,\sqrt{\frac{r_i^2}{16\,\pi\,\omega\,\omega_L\,r_H\,D}}\,\left(\frac{4
 \,\omega}{\omega_L\,r_H\,D}
\right)^{\frac{1}{6}}\,\,\tilde{Ai}(\xi;\zeta)\,\,;\label{AM3}\\
 \tilde{Ai}(\xi;\zeta) &=&
\Le\, \int_{C_1}\,+\,\int_{C_2}\,\Ra\,\frac{d\,\nu}{\sqrt{\nu}}\,\,e^{-\,\xi\,\nu
  \,\,+\,\,\frac{\nu^3}{3}\,\,-\,\,\frac{\zeta^2}{\nu}}\,\,;\label{AM4}\\
\xi &=& \left( \frac{D \,\omega_L\,r_H}{4 \,\omega}
\right)^{-\frac{1}{3}}\,
\{\frac{r_f\,+\,r_i}{2}\,-\,\frac{\omega_L\,r_H}{\omega}\,\}\,\,;\label{AM5}\\
\zeta&=& \frac{\delta r}{2}\,\left(\frac{\omega}{2\,D\,\omega_L \,r_H}
\right)^{\frac{1}{3}}\,
\mbox{with}\,\,\delta r = r_f \,-\,r_i\,\,;\label{AM6}
\een

Note, that the right dimension of solution of \eq{GBFKL}
must be $\,g(\omega,r_f, r_i)\,/\,\,(q_f\,q_i)\,\,$, but this fact is not important for our 
later consideration, so we will proceed with the dimensionless function \eq{AM3}.
Once $g(\omega,r_f, r_i)$ is known, 
the asymptotic behavior of our Green's  function $G(Y;r_f,r_i)$ 
can be obtained from  \eq{ME2}.  However, in dealing
with the general expression ( see
\eq{AM3} -
\eq{AM6} ) we have to
take into account that the limit depends on the virtualities $r_f$ and $r_i$
and, in particular, 
the value  $\tilde{\omega}$.
Further, ultimately, we need also to take into account the soft
Pomeron intercept. For different relations between these two values, different region in the
$\omega$ integral can dominate, leading
to different asymptotic behavior for the Green's function.
We start the search for the resulting singularities by considering
various situations where different regions  of   $\omega$  dominate the integral in \eq{ME2}.

 We first consider the case when $\omega\,>>\,\tilde{\omega}$.
In this region both $\xi $ ( \eq{AM5}) and $\zeta$ (\eq{AM6}) are large, and
therefore the main contribution in this integral
comes from the region of small $\nu$. Neglecting the $\nu^3$ term in the
exponent of the integrand, we obtain
\be\label{ORIG1}
\tilde{Ai}(\xi;\zeta)\,\,\propto\,\,
\int_{0}^{\infty}\,\frac{d\nu}{\sqrt{\nu}}\,
e^{-\nu\,\xi\,-\,
\frac{1}{\nu}\,\zeta^2}\,\,
\ee
We will see later, that omitted $e^{\nu^3\,/3\,}$ term is indeed small.
For such an integral the integration over two different contours in \eq{G1} 
give the same answer, since the integrand has no singularities in the
right semi plane and both integrals could be taken along the real axis.
The resulting integral can be evaluated explicitly, leading to:
\be\label{ORIG2}
\tilde{Ai}(r_i, r_f)\,\,\propto\,\,
2\,K_{-1/2}\Le\,2(\delta r/2)\sqrt{\frac{\omega\,-\,\tilde{\omega}}
{\tilde{\omega}\,D}}\,\Ra\,
\Le\,\frac{\zeta}{\xi^{1/2}}\,\Ra^{1/2}\,
\ee
where $K_{\nu}(z)$ is the modified (McDonald) Bessel function.
Since \eq{ORIG1} does not depend on the sign of $\delta r$, 
we can therefore write the answer as
\be\label{ORIG4}
g(\omega, r_f\,,r_i\,)\,=\,G^{1}_{\omega}(r_f\,, r_i\,)\,\,=\,\,
\frac{\,r_i\,}{\,2\,\tilde{r}^{1/2}}\,\,
\sqrt{\frac{1}{\omega\,\omega_L\,r_H\,D}}\,\,\Le\,
e^{-\,2\,(\frac{\delta\,r}{2})
\sqrt{\frac{\omega\,-\,\tilde{\omega}}{D\,\tilde{\omega}}}}\,
\theta(\delta\,r)\,+\,
e^{\,2\,(\frac{\delta\,r}{2})
\sqrt{\frac{\omega\,-\,\tilde{\omega}}{D\,\tilde{\omega}}}}\,
\theta(-\delta\,r)\,\Ra\,\,. 
\ee
Instead of integrating over $\delta r$, we could simply integrate over 
the following function for positive values of $\delta r$:
\be\label{ORIG555}
G^{1}_{\omega}(r_f\,, r_i\,)\,\,=\,\,
\frac{\,r_i\,}{\,\tilde{r}^{1/2}}\,\,
\sqrt{\frac{1}{\omega\,\omega_L\,r_H\,D}}\,\,
e^{-\,2\,(\frac{\delta\,r}{2})
\sqrt{\frac{\omega\,-\,\tilde{\omega}}{D\,\tilde{\omega}}}}\,\,,
\ee
where $\tilde{r}\,=\,\frac{r_i\,+\,r_f}{2}\,-
\,\frac{\omega_{L}\,r_{H}}{\omega}$.
In this region,  $\delta\,r$ and  $\nu $ are 
small,
$\delta\,r\,\sim\,\sqrt{\frac{D\,\tilde{\omega}}{\omega\,-\,\tilde{\omega}}}\,<<\,1$
and 
$\nu\,\propto\,\frac{\zeta}{\xi^{1/2}}\,\sim\,
\Le\frac{\tilde{\omega}}{32\,\omega}\Ra^{1/3}\,<<\,1$.
Therefore, we can trust this solution so long that the
following condition is satisfied:
\be\label{FCON}
\frac{\tilde{\omega}}{\,\omega}\,\leq\,1\,\,.
\ee
Indeed, even  when 
$\frac{\tilde{\omega}}{\,\omega}\,=\,1$, the  $\,\nu\,$
is still small, $\nu\,\sim\,
\Le\frac{\,1}{32\,}\Ra^{1/3}\,<<\,1\,\,.$
The magnitude of the  omitted term in our solution is also  small, namely,
$e^{\frac{1}{96}\,\frac{D\,\tilde{\omega}}{\omega\,(\frac{r_f\,+\,r_i}{2})^2}}$,
and we have for the Green's function in this region:
\be\label{ORIG5}
G^{1}_{\omega}(r_f, r_i)\,\,=\,\,
\frac{\,r_i\,}{\,\tilde{r}^{1/2}}\,\,
\sqrt{\frac{1}{\omega\,\omega_L\,r_H\,D}}\,\,
e^{-\,\delta\,r\,\sqrt{\frac{\omega\,-\,\tilde{\omega}}{D\,\tilde{\omega}}}
+\frac{1}{96}\,\frac{D\,\tilde{\omega}}{\omega\,(\frac{r_f\,+\,r_i}{2})^2}}\,\,,
\ee
This equation gives the asymptotic answer for the Green's function
for the case $\omega\,>>\,\tilde{\omega}$, in the  so called diffusion
approximation for the BFKL kernel.
The Mellin transform of \eq{ORIG5} gives:

\be\label{NAN}
G^{1}(Y;r_f\,, r_i\,)\,=\,\sqrt{\frac{1}{2\,\pi\,Y\,\tilde{\omega}\,D}}\,
e^{\,-\,\frac{(\delta r)^2}{4\,Y\,\tilde{\omega}\,D}}\,\,\,
\ee
( see calculations in Appendix A ).

\begin{figure}
\begin{center}
\epsfxsize=12cm
\leavevmode 
\hbox{ \epsffile{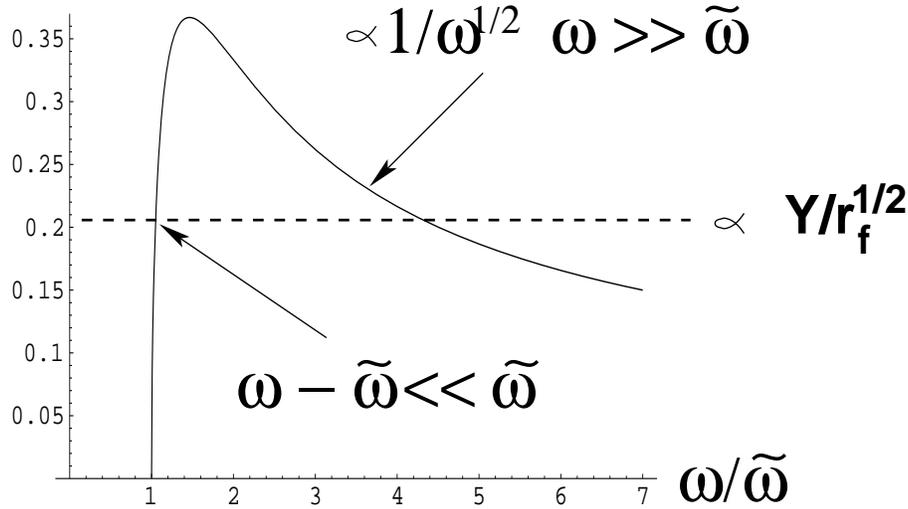}}
\end{center}
\caption{The positions of the saddle points in $\omega$ for $G(y;r_f,f_f)$ 
for $\omega > \tilde{\omega}$.}
\label{spom}
\end{figure}

There is another contribution in the limit of $\omega\,>>\tilde{\omega}$
coming from  the large $\nu$ region.
Indeed, it is easy to see from \eq{AM4}, that there is another saddle point, 
\be \label{SP1}
\nu^2_{SP}\,\,-\xi\, \simeq \, 0\,.
\ee
In the region $\xi\,>>\,1 $ we have
\be\label{SAD10}
\nu_{SP} \,=\,\,\sqrt{\xi}\, = \, 
\,r^{\frac{1}{2}}_f\frac{\sqrt{\omega \,-\,\tilde{\omega}}}{\omega^{\frac{1}{3}}}\,
( \omega_L\,r_H\,D/4)^{-\frac{1}{6}}\,>>1.
\ee
This saddle point corresponds  to the second crossing point
of Fig.~\ref{spom}. This is a solution  for the  $\omega > \tilde{\omega}$.

However, it turns out that in this regime,  $f^{-}_{SP}\, \gg \,1$ 
while $ f^{+}_{SP}$ is small ($ f^{+}_{SP} \ll \,1$).
It follows that we can not use the diffusion approach for the BFKL equation
to obtain an estimate of this contribution. In this case
\eq{AM3}-\eq{AM4} are no longer valid, and we have to take the full
kernel of \eq{OMEGA} into consideration. The integration over this full
 kernel is
performed along the real axis and instead of 
two solutions \eq{SOLME} and \eq{SOLME1}
now we have one.
Going back to the variables $f^+ $ and $f^-$ and considering
$f^{+}/f^{-}\,<<\,1\,$, one can obtain the following equation for the
saddle point:
\be \label{SPOM5}
\,r_f \,\,- \,\frac{2\,r_H}{\omega}\,\frac{\as(r_H)\,N_c}{\pi} \,
\frac{1}{1\,-\,f^{-}_{SP}}\,=\,\,0
\ee
 \eq{SPOM5} stems from the fact that
in the region of large $\omega$, $f^-$ is rather large. For
large $f^-$  we can replace $\omega(f^-)$ by $\omega(f^-)
\,\rightarrow\,
\frac{\as(r_H)\,N_c}{\pi}\,\frac{2}{1\,-\,f^{-}_{SP}}$. Thus we have
\be \label{SPOM6}
1 - f^-_{SP}\,\,=\,\,\frac{2 \tilde{\omega}}{\omega}\,\,\ll\,\,1\,.
\ee
We obtain for the  Green's function in this region of $\omega$:
\be\label{SPOM77}
G^{2}_{\omega}(r_i, r_f)\,\sim\,\sqrt{\tilde{\omega}}\,
e^{-\frac{r_i\,+\,r_f}{2}\,-
\frac{2\, \as(r_H)\,N_c\,r_H}{\pi\,\omega}\,
\,\ln(\frac{\tilde{\omega}}{\omega})\,\,}
\ee
For the $\omega$-saddle point, we have 
\be \label{SPOM7}
Y\,\,+\,\,\frac{2\, \as(r_H)
 \,N_c\,r_H}{\pi\,\omega^2_{SP}}\,\,
\ln(\frac{\tilde{\omega}}{\omega_{SP}})\,\,=\,\,0\,\,,
\ee
which leads to 
\be \label{SPOM8}
\omega_{SP}\,\, =\,\left(\frac{2\,\as(r_H)\,N_c\,r_H}{\pi\,\,Y}\, 
\ln\frac{2\,\as(r_H) N_c\,r_H}
{Y \pi\,\tilde{\omega}^2}\right)^{\frac{1}{2}}\,.
\ee
Using \eq{SPOM8}, we arrive at 
\be \label{SPOM9}
G^{2}(Y;r_f,r_f)\,\,\sim\,\,e^{2\, \sqrt{\frac{2\,\as(r_H)
\,N_c\,r_H}{\pi\,Y}\,\ln\frac{2\,\as(r_H) N_c\,r_H}
{Y \pi\,\tilde{\omega}^2}}}\,\,.
\ee 
This contribution has been missed in Refs.\cite{EL1,CTM}.
Nevertheless, comparing  Green's functions of \eq{ORIG5} and
 \eq{SPOM77},
we see,  in the limit of small $\tilde{\omega}$,
 the contribution of \eq{ORIG5}
is larger than the contribution of \eq{SPOM77}:
\be
\frac{G^{2}_{\omega}(r_i, r_f)\,}{G^{1}_{\omega}(r_i, r_f)\,}\,\sim\,
e^{-\frac{r_i\,+\,r_f}{2}\,}\,\,.
\ee

We next consider the limit where $\omega$  is close to 
$\tilde{\omega}$: $\omega\,=\,\tilde{\omega}\,+\,\delta\omega$,
 i.e., $\delta\omega\,/\,\tilde{\omega}\,<<1\,$.
In this case $\xi$ is small, $\xi\,<<\,1$, 
indeed, $\xi\,\sim\,\frac{\omega-\tilde{\omega}}{\omega}\,\propto\,
\frac{\delta\omega}{\tilde{\omega}}\,<<1\,$,
and for small $\xi$ we have  $\nu\sim\,1$.
We also take $\delta\,r$ in the limit $\delta\,r\,<<\,(r_H\,D)^{1/3}$
and, therefore, we neglect the $\zeta^2\,/\nu\,$ term in our equation.
We now need to consider both solutions of \eq{MEBFKL}
together, sum of \eq{SOLME} and \eq{SOLME1}. 
Due to the integration along the rays $\pi/3$ and $-\pi/3$ we have :
\be\label{EXAMP}
g(\omega,r_f, r_f)\,\,\propto\,
e^{i\pi/6}\,\int_{0}^{\infty}\,\frac{d\,\nu}{\sqrt{\nu}}\,\,
e^{-\,\xi\,\nu\,e^{i\pi/3}\,\,-\,\,\frac{\nu^3}{3}\,}\,+\,
e^{-i\pi/6}\,\int_{0}^{\infty}\,\frac{d\,\nu}{\sqrt{\nu}}\,\,
e^{-\,\xi\,\nu\,e^{-i\pi/3}\,\,-\,\,\frac{\nu^3}{3}\,}\,
\ee

and finally we obtain 
( see calculations in Appendix A ): 

\be\label{EQU2}
g(\omega,r_f, r_f)\,\,\propto\,
\frac{\Gamma(\frac{1}{6})}{3^{2/3}}\,-\,
\Le\frac{4\omega}{D\omega_L\,\,r_H}\Ra^{2/3}\,
\frac{(\frac{r_i\,+\,r_f}{2})^2}{8\,\omega^2}\,(\delta\omega)^{2}\,
\frac{\Gamma(\frac{5}{6})}{3^{4/3}}\,.
\ee

The Mellin transform of \eq{EQU2}, see Appendix A,  gives:

\be\label{EQU6}
G(Y;r_f\,,r_i\,)\,=\,\
\sqrt{\,\frac{3\,r_i}{8\,\pi \,Y\,\omega_L\,r_H\,D}}\,\,
e^{ \tilde{\omega}\,Y\,\,-\,\,\frac{2}{9\,\sqrt{3}}
\frac{D\,( \,\tilde{\omega}\,Y\,)^3}{(r_f\,+\,r_i)^2\,/4} }\,.
\ee

The region of applicability of this saddle point analysis stems from condition
that $f^+$ and $f^-$ should  be small in order to use the expansion of \eq{OMEGAD}
and from $\omega_{SP} < \tilde{\omega}$. It is easy to see that these
conditions restrict the value of rapidity $Y$ \cite{EL,KM,ABB,EL1,CTM}

\be \label{SPOM4}
0 \,<\,Y\,<\, \frac{r^2_f}{\omega_L\,r_H}\,\,.
\ee

Now, if we consider together the asymptotic behavior of the Green's
function given by \eq{NAN} and \eq{EQU6},
we see that in a large range of rapidity, given by \eq{SPOM4},
the contribution of  \eq{EQU6}
is much larger than contribution 
\eq{NAN}. Therefore, the main contribution to the 
Green's function gives the region 
of such $\omega$,   where $\omega\,\sim\,\tilde{\omega}$.

  In the region of small $\omega \,<\,\tilde{\omega}$ where $\xi $ is negative
(see   \eq{AM5} ),  the saddle point of $\nu$ is equal to
\be \label{SPOM10}
\nu_{SP}\,\,=\,\,\pm \,i\,\left(\frac{\omega_L\,r_H}{D\,\omega}
 \right)^{\frac{1}{3}}\,\,.
\ee
Performing calculations for the saddle point we arrive at
\be \label{SPOM11}
\omega_{SP}\,\,=\,\,\pm\,\,\Le 1 - i\Ra\,\,\sqrt{
\frac{2\,\omega_L\,r_H}{D\,3\,Y}} \,\,.
\ee
The Green's function is \cite{CTM}
\be \label{SPOM12}
G(Y;r_f,r_f)\,\,\sim\,\,e^{2\,\,\sqrt{\frac{\omega_L\,r_H\,Y}{3\,D}}}\,\,\cos
\,\sqrt{\frac{\omega_L\,r_H\,Y}{3\,D}}
\,\,.
\ee
It turns out that restrictions $f^{+}_{SP} <1$ and $f^{-}_{SP} < 1$ can be satisfied,
but only within numerical accuracy, namely, $f^{-}_{SP} < 1/D$. We will see
later that $D$ is rather large.

\subsection{Soft to BFKL Pomeron transition }

\begin{figure}
\begin{center}
\epsfxsize=8cm
\epsfysize=16cm
\leavevmode 
\hbox{ \epsffile{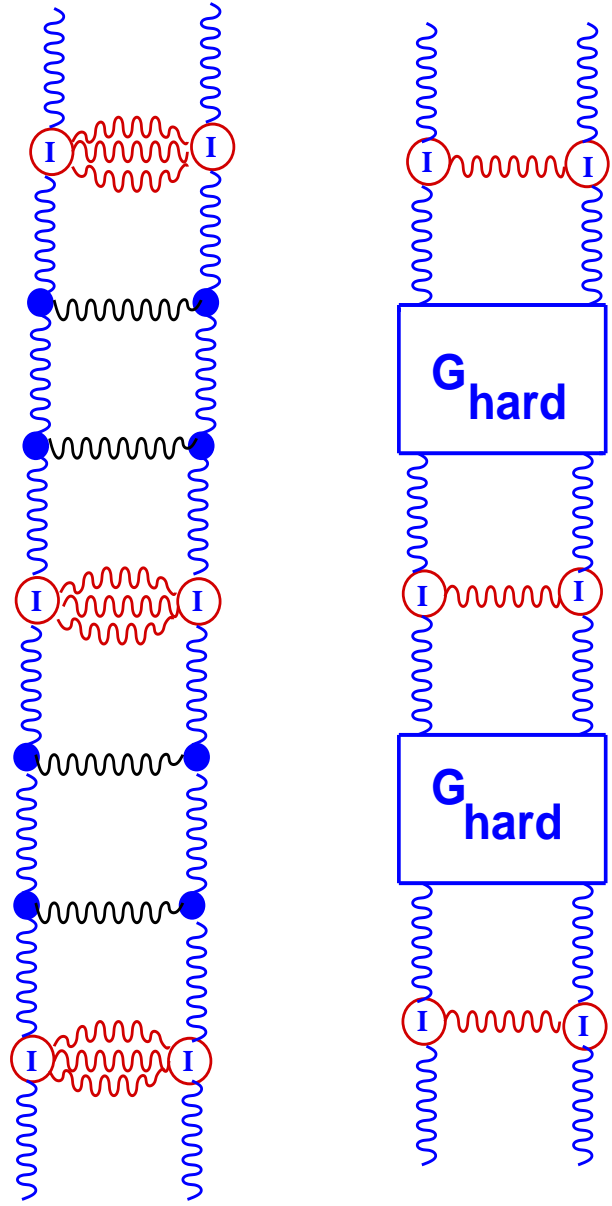}}
\end{center}
\caption{The diagrams and the graphic form of the equation for soft to BFKL Pomeron transition.}
\label{S-H}
 \end{figure}

The main idea of the paper is shown in Fig.~\ref{S-H}. 
Generally speaking the sum of diagrams in Fig.~\ref{S-H} can be found as a solution to 
the following equation:
\be \label{SHT01}
G^{S-H}_{\omega}(r_f,r_i)\,\,\,=\,\,\,D(\omega;r_f,r_i)\,\,\,\,+
\,\,D\,\bigotimes\,\left(K_S + K_H\right)\,\bigotimes\,G^{S-H}_{\omega}\,\,.
\ee
where sign $\bigotimes$ stands for all needed integrations and $D(\omega;r_f,r_i)$ 
denotes the Green's function of two gluon exchange.
Using the factorization properties of soft Pomeron kernel we can solve this equation,
the derivation of the  solution is given in Appendix B, and it reads :

\be\label{MASTER} 
G^{S-H}_{\omega}(r_f,r_i)\,\,=\,G^{H}_{\omega}(r_f,r_i)\,+
\,\frac{\tilde{G}^{H}_{\omega}(r_f,r_i)}{1\,\,-\,\,\Delta_S
\int \,d r'\,\,\int \,d r' \,\phi(r')\,\,G^H_{\omega}(r',
r'')\,\,\phi(r'')}\,\,,
\ee
where
\be\label{MASTER_11}
\tilde{G}^{H}_{\omega}(r_f,r_i)\,=\,
\Delta_S\,\int \,d r'' \,\phi(r'')\,\,G^H_{\omega}(r_f,
r'')\,\int \,d r' \,\phi(r')\,\,G^H_{\omega}(r',r_i)\,\,.
\ee

It is easy to see from \eq{MASTER_11}, that when $\,K_S= 0\,$, \eq{MASTER} describes only the 
``hard'' Pomeron. In the case, where there is no pQCD emission, when  $\,\,G^H = D\,\,$, 
\eq{MASTER} has a form:

\be \label{SHT111}
G^{S-H}_{\omega}(r_f,r_i)\,\,=\,\,
D(\omega;r_f,r_i)\,+\,\Delta_S\,
\,\frac{\int \,d r'' \,\phi(r'')\,\,D(\omega;r_f,r_i)\,\int \,d r' \,\phi(r')\,
\,D(\omega;r_f,r_i)\,}{1\,\,-\,\,\Delta_S
\int \,d r'\,\,\int \,d r' \,\phi(r')\,D(\omega;r_f,r_i)\,\,\phi(r'')}\,\,.
\ee
Since $\,\,D(\omega;r_f,r_i)\,\,=\,\,\delta(r_f - r_i)/\omega\,\,\,$, \eq{SHT111} leads to
\be \label{SHT112} 
G^{S-H}_{\omega}(r_f,r_i)\,\,=\,\frac{\delta(r_f - r_i)}{\omega}\,+\,
\frac{1}{\omega}\,\frac{K_{S}(r_f,r_i)}{\omega \,\,-\,\,\Delta_S}\,\,,
\ee 
or, in other words, we only have the soft Pomeron in this case.

\eq{MASTER} has poles which are the zeros of the denominator, namely,

\be \label{SHT2}
1\,\,-\,\,\Delta_S
\int \,d r'\,\,\int \,d r'' \,\phi(r')\,\,G^H_{\omega}(r',
r'')\,\,\phi(r'')\,\,\,=\,\,\,0\,\,.
\ee 

For $\omega \,>>\,\tilde{\omega}$
the solution of \eq{SHT2} is (see Appendix C )
\be\label{EXP2}
\omega\,=\,\Delta_{S-H}\,=\,\Delta_{S}\,+\,\tilde{\omega}_{S}\,+\,
\frac{D\,\tilde{\omega}_S\,\Delta_S}{96\,r_{S}^2\,(\tilde{\omega}_S\,+\,
\Delta_S)}\,.
\ee

Therefore,   the intercept of the resulting pole is 
simply $\Delta_{S-H}\,=\Delta_S\,$ for
$\tilde{\omega}_S\,\rightarrow\,0\,$.
This result is expected since for $\tilde{\omega}_S
\,=\,0$, there is no perturbative emission of gluons.
It should be stressed,  that this answer is obtained with the 
restriction,
\be\label{MARE}
\omega\,\geq\,\tilde{\omega}_S\,\,,
\ee
see also \eq{FCON}.

It is easy to see, that this restriction is satisfied when $\tilde{\omega}_S$
is small. This happens in two cases:
\begin{itemize}
\item when $\bar{\alpha}_S $ is small.
In this case $\tilde{\omega}$ is also small and, therefore, from
\eq{MARE}
we have:
\be\label{USL1}
\Delta_S\,>>\tilde{\omega}_S\,\sim\,\tilde{\omega}\,\,.
\ee
We can make a  simple numerical estimate: for 
$\bar{\alpha}_S\,\approx\,0.1$
we have $\omega_L\,\approx\,0.18\,\,(\tilde{\omega}_S\,\approx\,0.1\,)$ 
(see next section), so
taking $\omega=\Delta_{S-H}\,\approx\,0.2$ we  obtain
$\Delta_{S}\sim\,0.1$, where 
$\Delta_S\,>\,\tilde{\omega}_S\,\sim\,\tilde{\omega}$.
So, one can  see that restriction \eq{USL1} is satisfied for
$\bar{\alpha}_S\,\leq\,0.1$;  

\item when $\tilde{\omega}_S$ is small due to
possible large $r_S$, namely,
\be\label{USL2}
r_S\,\geq\,\frac{\omega_L\,\,r_H}
{\Delta_{S-H}\,}\,\,.
\ee
\end{itemize}

\begin{figure}
\begin{center}
\epsfxsize=12cm
\leavevmode
\hbox{ \epsffile{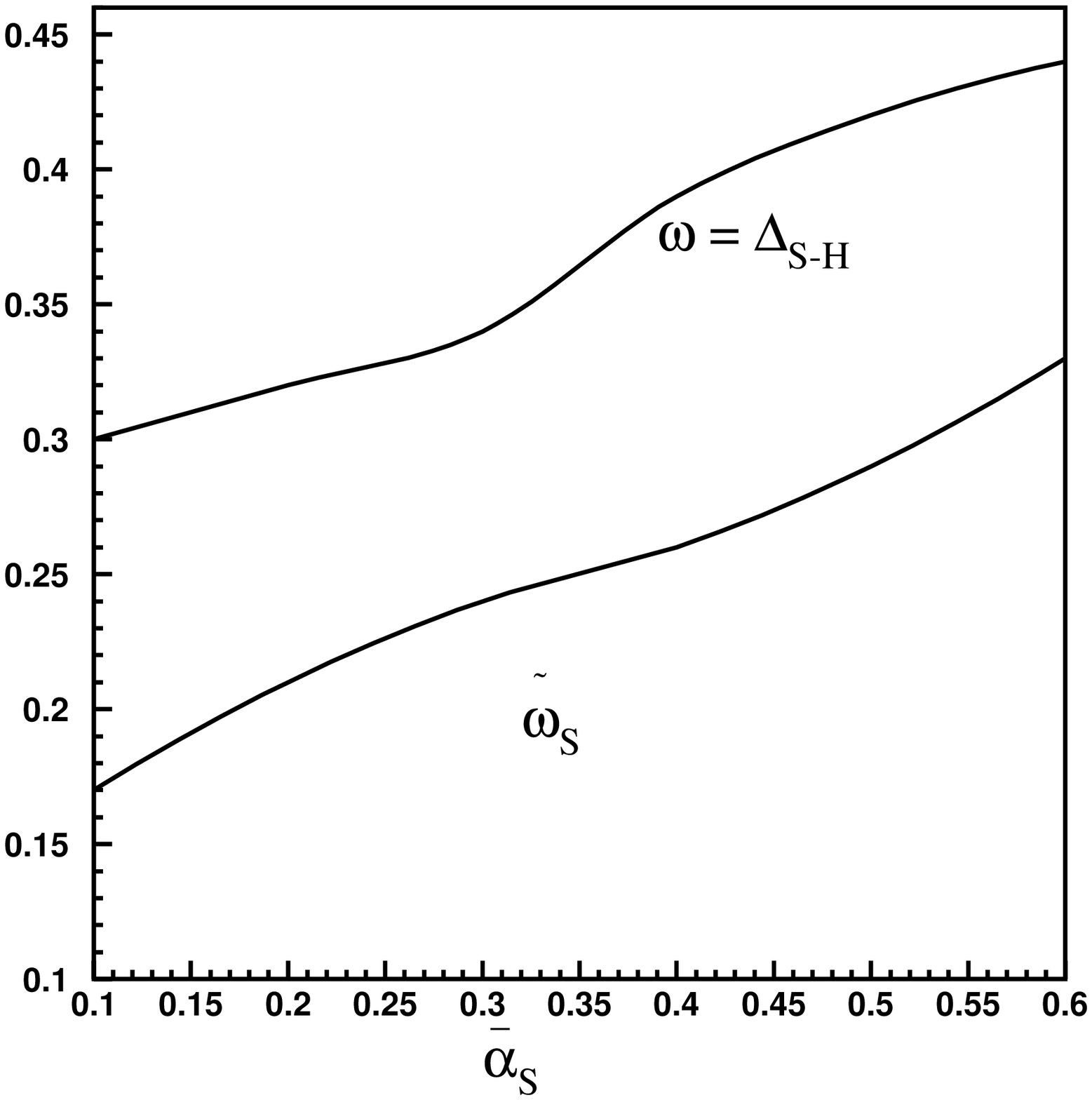 }}
\end{center}
\centerline{}
\caption{Solutions for $\Delta_{S-H}$ and $\tilde{\omega}_S$ 
as functions of $\,\bar{\alpha}_S\,$ at $\Delta_s = 0.1$.}
 \label{solut}
\end{figure}

We next consider  another region of $\omega$. We assume that
the position of the resulting pole $\omega \,=\,\Delta_{S-H}$ may be    
close to
$\tilde{\omega}_S\,$, $
\,\Delta_{S-H}\,=\,\tilde{\omega}_S\,+\,\delta\omega\,$,
where for  $\delta\omega$ there are two asymptotic limits. 
The first is the small $\delta\omega\,\,$, 
$\frac{\delta\omega\,}{\,\tilde{\omega}_S}\,<<\,\,1\,$, and 
the second for the $\delta\omega\,$ of the order of $\tilde{\omega}_S\,\,$,
$\frac{\delta\omega\,}{\,\tilde{\omega}_S}\,\sim\,\,1\,\,$.
The solution of  \eq{SHT2} in this case is

\be\label{USL44}
\omega\,=\Delta_{S-H}\,=\,\tilde{\omega}_S\,+\,\delta\omega\,=\,
\frac{2\,\Delta_{S}^{3}\,r_S\,}{\tilde{\omega}_{S}^{2}\,D^{2}\,}\,
\Le\,\frac{\Gamma(\frac{1}{6})}{3^{2/3}}\,\Ra^{3}\,\,.
\ee
Fig.~\ref{solut} shows the numerical solution  to \eq{USL44}
with  $r_S=r^{mat}_{S}\,$ (see later \eq{MATCH}), and with
$\Delta_S\,=\,0.1\,$. We obtain that 
for $\Delta_s = 0.1$,
$\frac{\,\delta\omega\,}{\tilde{\omega}_S\,}\,\sim\,0.3\,-\,0.8$.

We can trust this solution only if $\omega $ is larger than $\,\tilde{\omega}_S\,$,
and this condition can be translated in the following inequality
\be\label{USL45}
\,\tilde{\omega}_S\,<\,
\frac{2\,\Delta_{S}^{3}\,r_S\,}{\tilde{\omega}_{S}^{2}\,D^{2}\,}\,
\Le\,\frac{\Gamma(\frac{1}{6})}{3^{2/3}}\,\Ra^{3}\,\,.
\ee
or in the restriction on the value of $\,\tilde{\omega}_S$
\be\label{EXP3}
\tilde \omega_S\,\leq \,\Delta_S\,\frac{r_{S}^{1/3}}{D^{2/3}}\,
\frac{2^{1/3}\,\Gamma(\frac{1}{6})}{3^{2/3}}\,\,.
\ee
Remembering that \eq{USL44} is valid only if $\tilde \omega_S
\,\geq\,\Delta_S\,$ we see that \eq{USL45} is a solution for $r_S$ which
satisfies the inequality
\be \label{INQ1}
\Delta_S\,\,\leq\,\,\tilde{\omega}_S\,\,\leq\,\,\Delta_S\,\frac{r_{S}^{1/3}}{D^{2/3}}\,
\frac{2^{1/3}\,\Gamma(\frac{1}{6})}{3^{2/3}}\,\,.
\ee
The solution of \eq{INQ1} defines the region of $r_S$ for which 
we can trust the approximate equation \eq{USL44}.

\begin{figure}
\begin{center}
\epsfxsize=12cm
\leavevmode
\hbox{ \epsffile{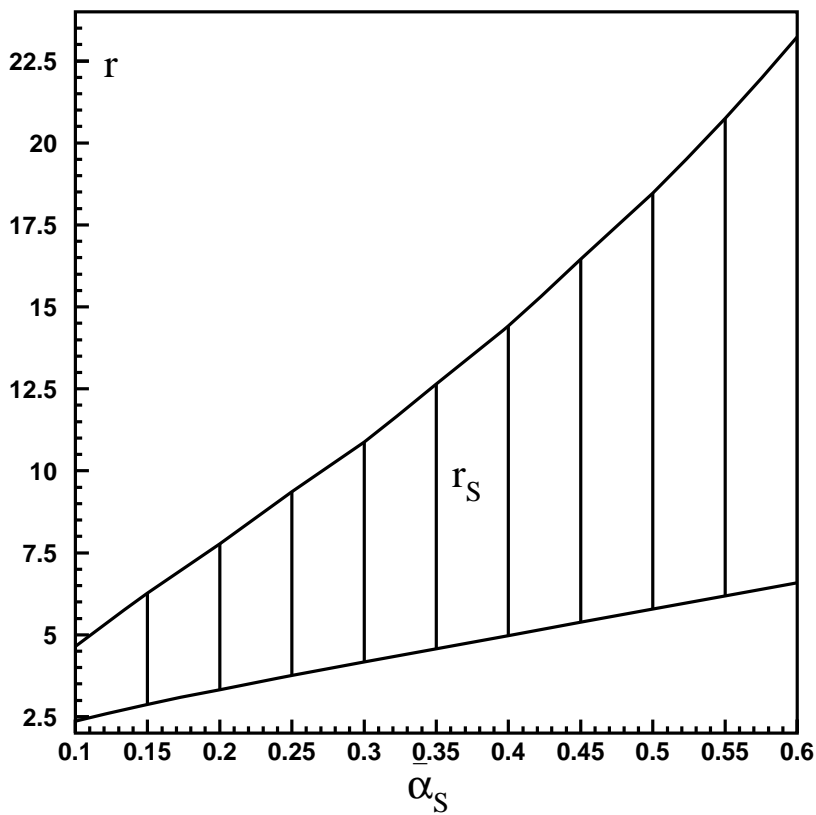 }}
\end{center}
\centerline{}
\caption{Permitted  $\,r_S\,$ as function of $\,\bar{\alpha}_S\,$ at 
$\Delta_S\,=\,0.1$. }
\label{r_alpha}
\end{figure}

\begin{figure}
\begin{center}
\epsfxsize=12cm
\leavevmode
\hbox{ \epsffile{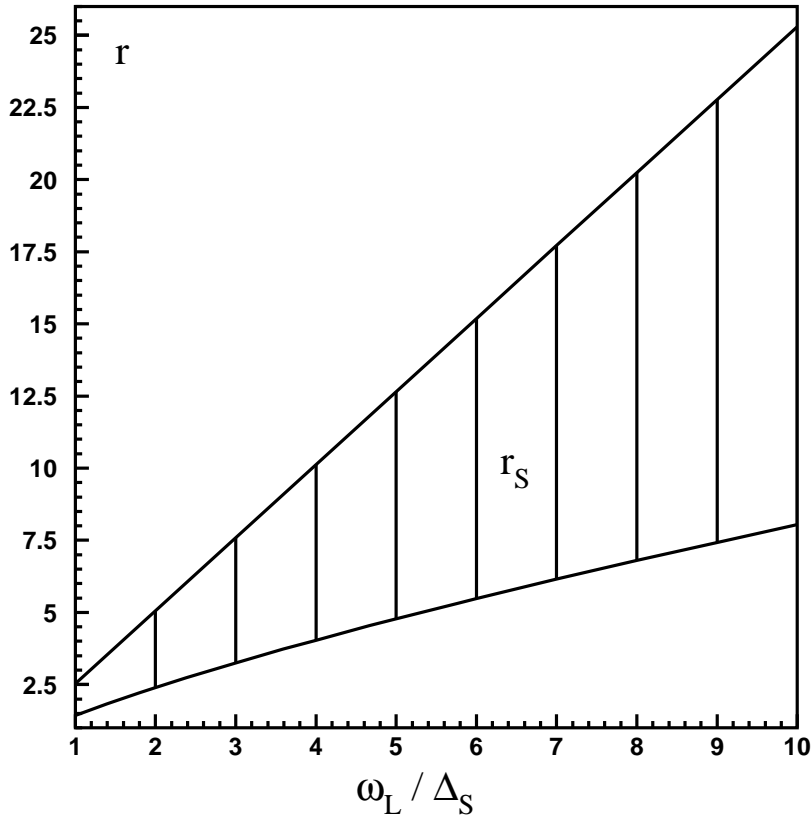 }}
\end{center}
\centerline{}
\caption{Permitted  $\,r_S\,$ as function of $\,\frac{\omega_L}{\Delta_S}\,$
at $D\,=\,3\,$. }
 \label{r_alpha_omega}
\end{figure}

 In Fig.~\ref{r_alpha}, we present  the solution for $\,r_S\,$ to \eq{INQ1} as a  
function of $\,\bar{\alpha}_S\,$ at fixed  $\Delta_S\,=\,0.1$. 
In  Fig.~\ref{r_alpha_omega}, the same solution is plotted versus
$\,\frac{\omega_L}{\Delta_S}\,$
for $D\,=\,3\,$.

\begin{figure}
\begin{center}
\epsfxsize=12cm
\leavevmode
\hbox{ \epsffile{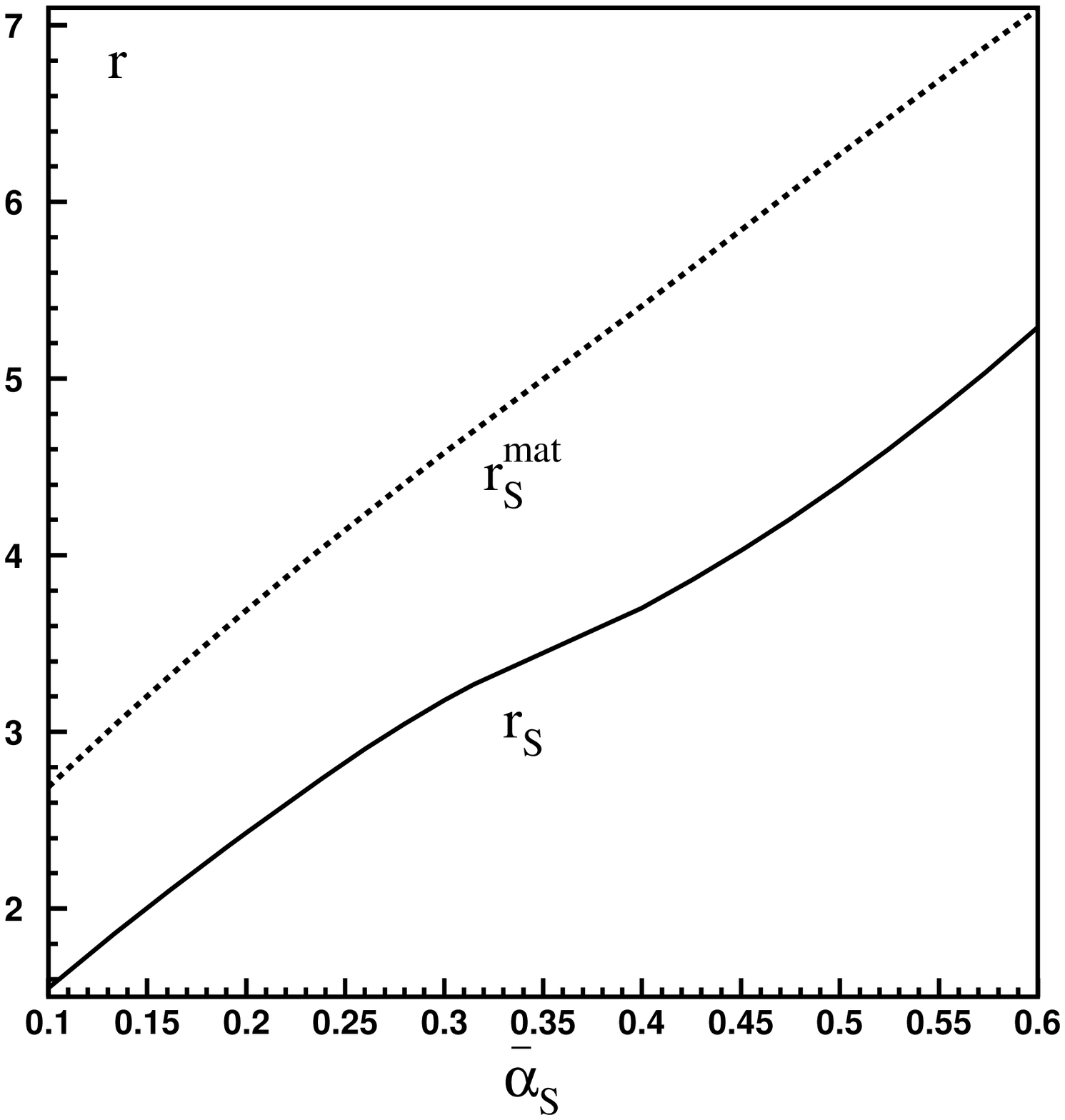}}
\end{center}
\centerline{Matching value $r_{S}^{mat}$ and boundary value $r_S$
as function of  $\,\bar{\alpha}_S\,$ at 
$\Delta_S\,=\,0.1$.}
\caption{ }
\label{r_S_mat}
\end{figure}

 The  matching between \eq{USL44} and \eq{EXP2} occurs at  
$r_S\,=\,r^{mat}_{S}\,$, for which we have 
\be \label{MATCH}
\Delta_{S-H}\,\,=\,\,\Delta_S \,+\,\tilde \omega_S \,+\,\frac{D\,\tilde
\omega_S \,\Delta_S}{96\,r^2_S\,(\tilde \omega_S + \Delta_S)}\,\,=\,\,
\frac{2\,\Delta_{S}^{3}\,r_{S}}{\tilde{\omega}^{2}_{S}\,D^{2}}\,
\Le\frac{\,\Gamma(\frac{1}{6})}{3^{2/3}}\Ra^{3}\,\,.
\ee
A numerical solution of this equation is presented
in  Fig.~\ref{r_S_mat}. This solution is the function of
$\,\bar{\alpha}_S\,$ with $\,\Delta_S  = 0.1\,.$ 
To justify this procedure, we have to show  that, for $r_S\,=\,r^{mat}_{S}\,$
both solutions, \eq{USL44} and \eq{EXP2}, are valid.
It is easy to see from Fig.~\ref{r_alpha}
that    $\,r^{mat}_{S}\,$ calculated 
from \eq{MATCH} for any value of
$\,\bar{\alpha}_S\,$,  
is  in region where
\eq{USL44} has solution.  
From Fig.~\ref{solut} we also obtain, that for given $\,r_S\,$
the ratio
$\frac{\,\delta\omega\,}{\tilde{\omega}_S\,}\,$ is
$\frac{\,\delta\omega\,}{\tilde{\omega}_S\,}\,\sim\,0.3\,-\,0.8\,\,$,
and these values satisfy the condition of \eq{conver},
which defines the applicability of this solution.
The  solution  \eq{USL44}
is valid for  $r_S\,=\,r^{mat}_{S}\,$.
We  next  consider the solution given by \eq{EXP2}. 
The region of applicability of this solution is defined by \eq{USL2}:
\be
r_S\,\geq\,\frac{\omega_L\,r_H}{2\,\Delta_{S-H}}
\ee
The boundary value of $r_S$, which is 
$r_S\,=\,\frac{\omega_L\,r_H}{2\,\Delta_{S-H}}$,
is also  presented in the Fig.~\ref {r_S_mat}. Considering 
this picture,  we  see, 
that this line lies under the $r_{S}^{mat}$
defined by \eq{MATCH}. 
Remembering, that the ratio
$\frac{\,\delta\omega\,}{\tilde{\omega}_S\,}\,\sim\,0.3\,-\,0.8\,\,$
satisfies the \eq{FCON}, we obtain, 
that solution \eq{EXP2}
is also valid for the given values of $r_{S}^{mat}\,.$

So,  finally, we have two solutions, 
the first is \eq{EXP2} which is useful for $r_S$ that
lies between the lines of Fig.~\ref{r_S_mat},
and second is \eq{USL44}, which we use when  $r_S\,$ 
lies between the line for $r_{S}^{mat}$ from Fig.~\ref{r_S_mat} and upper bound 
of Fig.~\ref{r_alpha}.
In the following we will denote all  these solutions
by the  $\Delta_{S-H}$.

\section{NLO BFKL}

As have been discussed the position of the resulting singularity (pole)
crucially depends on the value of $\tilde\omega$ or $\omega_L$. 
 In this section we discuss the values of 
$\omega_L$ and $D$  in the NLO BFKL equation\cite{LF,CCNO}:
\be
\omega=\omega_L\,\Le 1+D\,f^2\Ra.
\ee
NLO BFKL equation leads to  the following values
of $\omega_L$ and $D$ (see Refs. \cite{EL1,CCNO,R}):
$$
\omega_L=\bar{\alpha}_S\Le q^2_0\Ra\,\Le 2.772-
18.3\,\bar{\alpha}_S\Ra,
$$
and
\be
D=\bar{\alpha}_S\Le q^2_0\Ra\,\Le 16.828-
322\,\bar{\alpha}_S\Ra\, ,
\ee
These values have a serious
pathology, as for 
$\bar{\alpha}_S\,>\,0.152\,$,$\Le q_0^2\,<\,30\,GeV^2 \Ra$, 
$\omega_L$ becomes negative and for 
$\bar{\alpha}_S\,\sim\,0.05\,,\Le q_0^2\,<\,3\times\,10^6\,GeV^2 \Ra$ 
$D$ also changes  sign. A solution to this problem 
was proposed by G.P.Salam \cite{GPS} and 
M.Ciafaloni, D.Colferai and G.P.Salam \cite{CCS}.
They suggested  a 
scheme to improve (resummation)  the NLO BFKL kernel.
Their resummation is based on the idea that the resumed NLO BFKL kernel
should reproduce the NLO DGLAP kernel in the region of large photon
virtualities.
 Indeed, when we start to  consider the NLO terms
\be
\alpha_s\,\Le \alpha_s\ln\,\frac{s}{s_0}\Ra^{n}
\ee
and change the energy  scale  in the DGLAP evolution equation,
namely, $\,s_0\,=
\,k^2_1\,\Le k_1^2\,\gg\,k_2^2\,\Ra\,$
to the energy  scale $\,s_0=k_1\,k_2\,$ typical for the BFKL approach, the 
double logarithmic terms  arise: 
\be
\Le \alpha_s\,\ln^2\,\frac{k_1^2}{k^2_2}\Ra^{n-m}
\,\Le \alpha_s\,\ln\,\frac{k_1^2}{k^2_2}\,\ln\,\frac{s}{k_1\,k_2}\Ra^{m}.
\ee
These terms, formally subleading, nevertheless may give
 significant corrections to the  NLO kernel. It was shown
\cite{GPS,CCS} that these terms are 
 responsible for the difficulties with
the NLO kernel. Therefore, the proposed  resummation of these double
logarithmic terms was done so, that 
the resummed kernel reproduces a correct
DGLAP limit for $\omega\,\,\pm\,1/2\,$ and 
for  a scale $\,s_0\,=\,k^2_1\,$. Using this
method of resummation we  can avoid the difficulties of the NLO BFKL
kernel.  This improved kernel,
reproduces correctly  the  exact
NLO BFKL kernel in the region of its applicability, but it turns out to be
free  from inconsistency of ''pure''
NLO BFKL kernel.  From the  proposed different schemes for the
improved kernel, see \cite{GPS}, we use scheme number four,
as this scheme and also scheme number three 
has a reasonable  behavior of second 
derivative of the kernel as a function of $\tilde{\alpha_s}$.
As the additional  resummation of the terms which are important for the 
second derivative of the kernel was made.

\begin{figure}
\begin{center}
\epsfxsize=12cm
\leavevmode
\hbox{ \epsffile{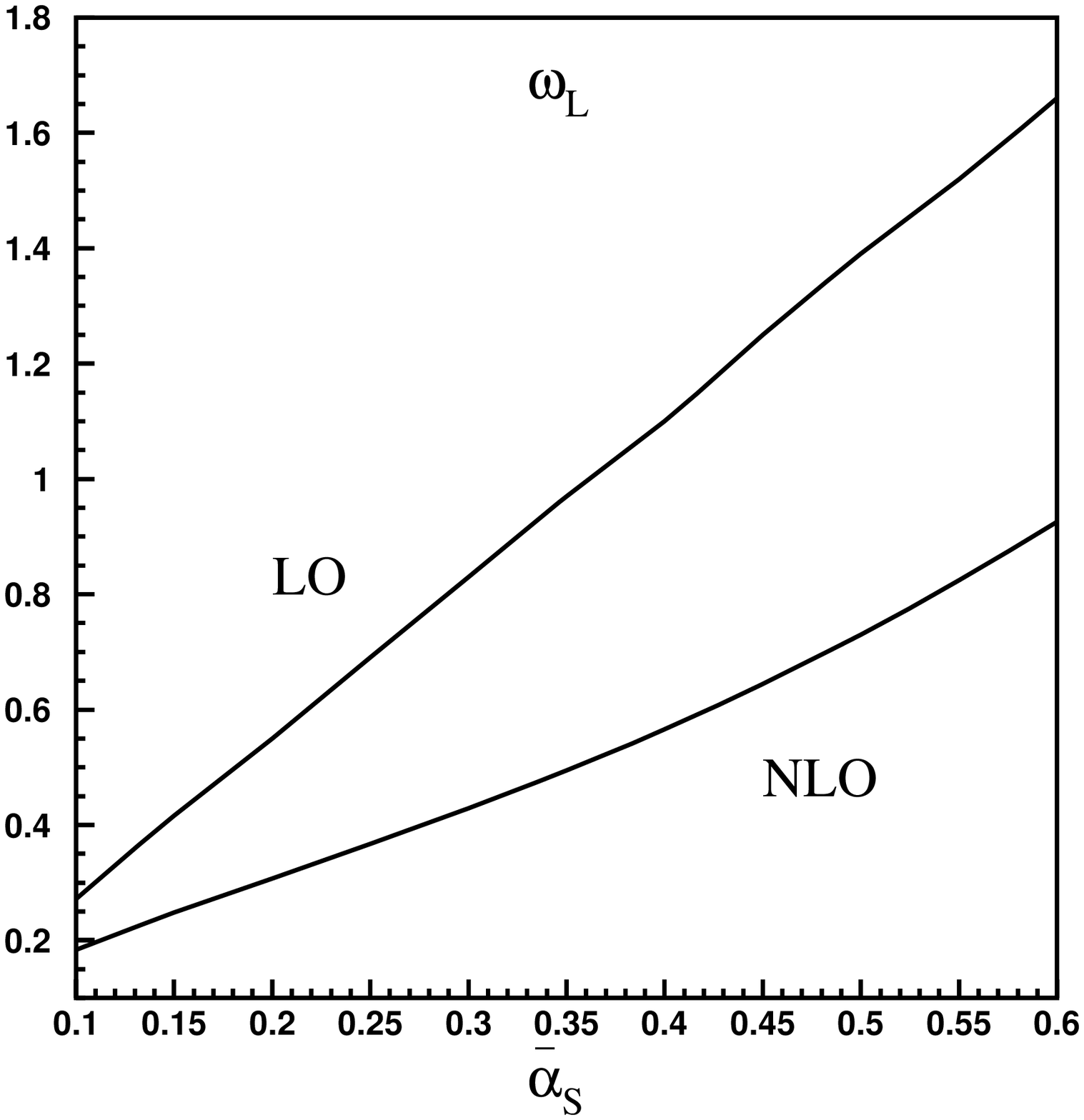 }}
\end{center}
\centerline{}
\caption{ }
 \label{BFKL-A}
\end{figure}

\begin{figure}
\begin{center}
\epsfxsize=12cm
\leavevmode
\hbox{ \epsffile{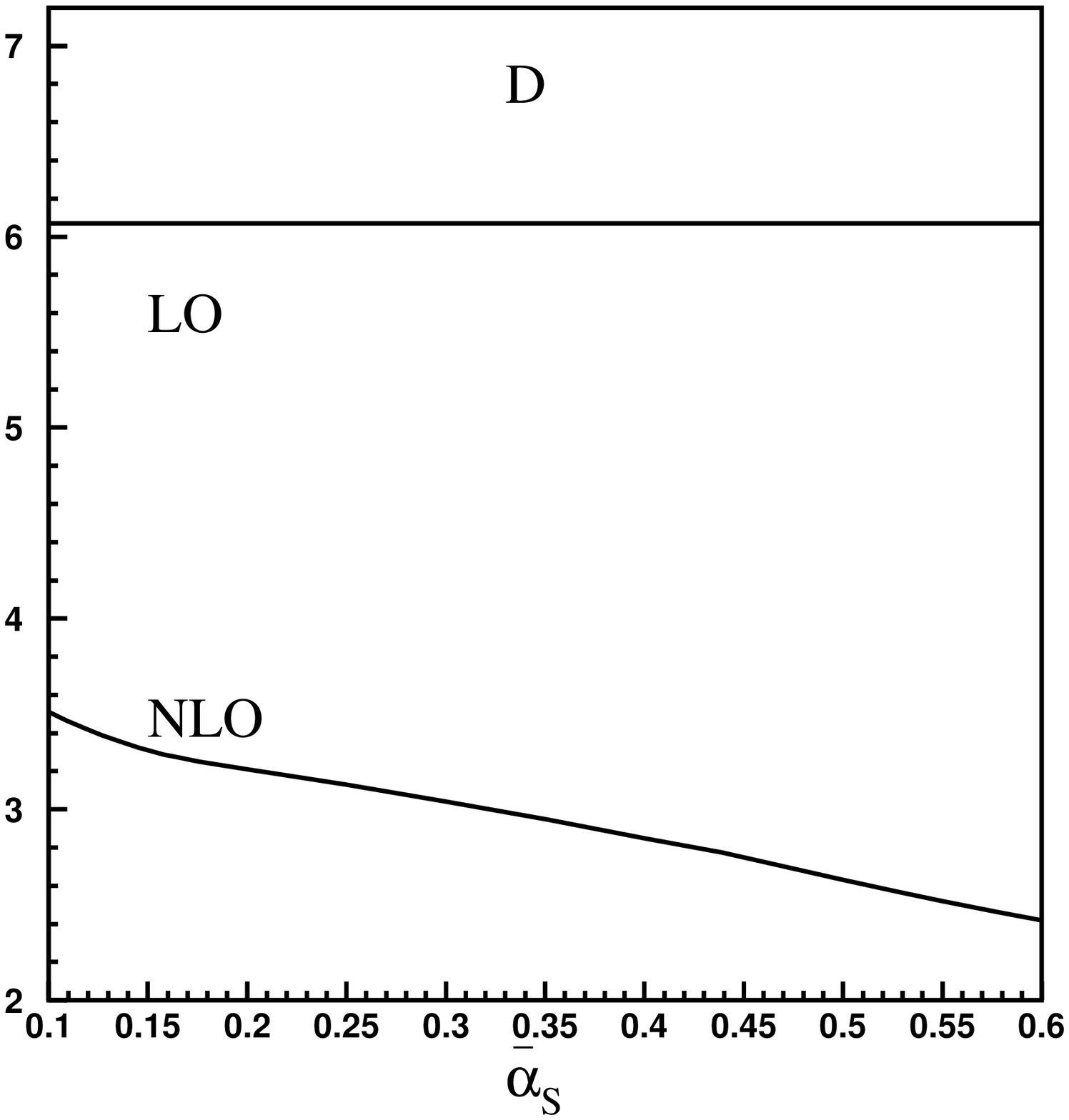 }}
\end{center}
\centerline{}
\caption{ }
 \label{BFKL-D}
\end{figure}

 So we have:
$$
\bar{\chi}\Le \gamma\,\omega\,\bar{\alpha}_s\Ra=
$$
$$
\bar{\alpha}_s\Le\chi_0\Le\gamma\Ra-\frac{1}{1/2-\gamma/2}-
\frac{1}{1/2+\gamma/2}+\frac{1}{1/2-\gamma/2+\omega/2+\bar{\alpha}_s\,B}+
\frac{1}{1/2+\gamma/2+\omega/2+\bar{\alpha}_s\,B}\Ra+
$$
$$
\bar{\alpha}_s^2\Le\chi_1\Le\gamma\Ra+
\Le B+\frac{1}{2}\,\chi_0\Le\gamma\Ra\,\Ra\Le
\frac{1}{\Le 1/2-\gamma/2\Ra^2}+\frac{1}{\Le 1/2+\gamma/2\Ra^2}\Ra\Ra+
$$
\be
\bar{\alpha}_s^2\Le
A^{'}\,\Le\frac{1}{1/2-\gamma/2}+\frac{1}{1/2+\gamma/2}-
\frac{1}{1/2-\gamma/2+\omega/2+\bar{\alpha}_s\,B}-
\frac{1}{1/2+\gamma/2+\omega/2+\bar{\alpha}_s\,B}\Ra\Ra\,. 
\ee
Here 
$$
\chi\Le \gamma,\,\omega,\,\bar{\alpha}_s\Ra=
\bar{\alpha}_s\,\chi_0\Le\gamma\Ra+
\bar{\alpha}_s^2\,\chi_1\Le\gamma\Ra 
$$ 
is 
usual BFKL kernel in the next-to-leading order,
and
\be
B=\frac{11}{8}-\frac{n_f}{12\,N_c}+\frac{n_f}{6\,N_c^3},
\ee
\be 
A^{'}=-\frac{1}{2}+\frac{5\,n_f}{18\,N_c}+\frac{13\,n_f}{36\,N_c^3}.
\ee
Solving the NLO BFKL equation with improved kernel
\be
\omega=\bar{\chi}\Le \gamma,\,\omega,\,\bar{\alpha}_s\Ra,
\ee
we can calculate our NLO
$\omega_L$ and $D$. 
The results are presented in  Fig.~\ref{BFKL-A} 
and Fig.~\ref{BFKL-D},
where we also plotted LO $\omega_L$ and $D$ .
These figures show that the resummed NLO BFKL kernel gives smaller values
both for $\omega_L$ and for $D$. From Fig.~\ref{BFKL-A} we obtain $\tilde
\omega \,\approx\,0.2\,$, which is too high to describe the experimental
data for soft processes. On the other hand, namely, this value of the
Pomeron intercept describes the inclusive data \cite{INCL}.

\section{Hard and soft Pomerons interface in DIS }

\begin{figure}
\begin{center}
\epsfxsize=12cm
\leavevmode
\hbox{ \epsffile{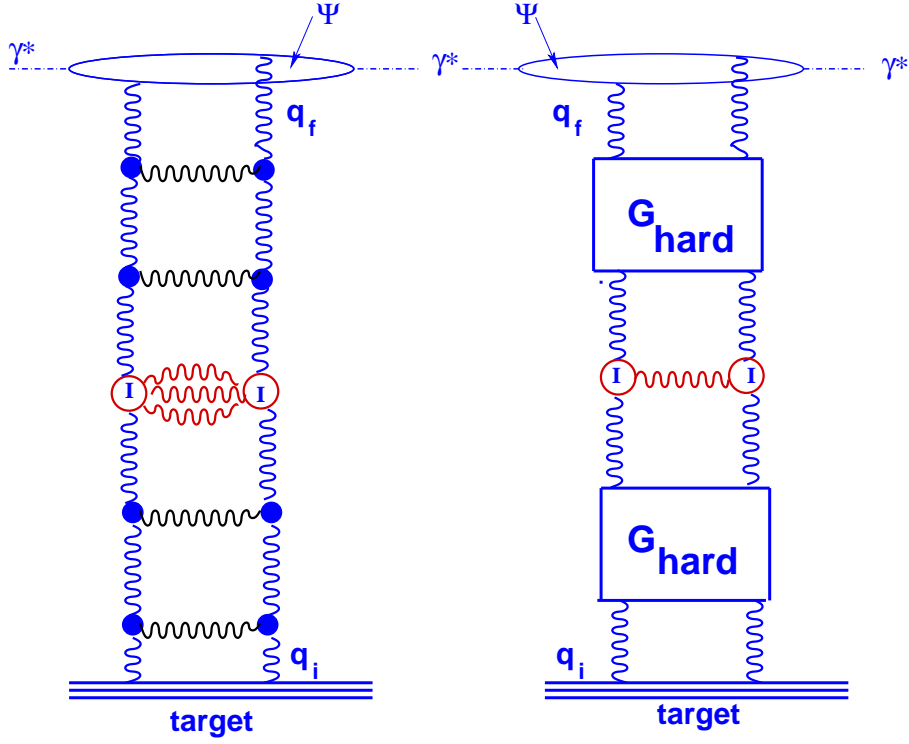 }}
\end{center}
\centerline{}
\caption{Diagrams and graphic form of contributions to deep inelastic 
processes.}
\label{DISSH}
\end{figure}

In this section we consider the deep inelastic processes in which we can 
observe our ``effective'' Pomeron considering the $Q^2$ 
- dependence of the cross section ( $Q^2$ is the photon virtuality) .
The diagrams that contribute to DIS are shown in Fig.~\ref{DISSH}.

The DIS total  cross section can be written in the form
\be \label{DIS1}
 \sigma_{tot}(\gamma^* p) \,\,=\,\,\int d r_f d r_i J^{\gamma^*}(Q^2; 
r_f)\,\,G^{DIS}(Y,r_f,r_i) \,\,J^{N}(r_H,r_i)
\,\,,
\ee
where $J^{\gamma^*}$ and $J^{N}$ are photon and nucleon impact factors, 
which can be easily calculated in the LO of perturbative QCD 
(see  Refs.\cite{IMF} and references therein). For example,
\be \label{DIS2}
J^{\gamma^*}(Q^2; r_f)\,\,=\,\,\int \,d^2 r_{\perp}\,\, \int^1_0\,\,d 
z \,|\,\Psi 
(Q^2; r_{\perp},z) |^2 \,\,\left( \,1\,\,-\,\,e^{i \vec{q}_f \cdot 
\vec{r}_{\perp}}\,\right)\,\,.
\ee
We introduce the Mellin transform of $\,G^{DIS}(Y,r_f,r_i)\,$ 
with respect to Y
\be\label{TO1}
G^{DIS}(\,Y\,,r_f\,,r_i\,)\,\,=\,\,\frac{1}{\,q_f\,q_i\,}
\int^{a+i\,\infty}_{a-i\,\infty}\,\frac{d\,\omega}{2\,\pi\,i}\,
\Le\,\frac{s}{q_f\,q_i}\,\Ra^{\,\omega\,}\,G^{DIS}_{\omega}(\,r_f\,,r_i\,).
\ee
$G^{DIS}_{\omega}$ is  sum of diagrams of Fig.~\ref{DISSH} and, therefore, it is equal to 
$G^{S-H}_{\omega}$ given by \eq{MASTER}:

\be \label{MASTER1}
G^{DIS}_{\omega}(\,r_f\,,r_i\,)=
\,G^{H}_{\omega}(r_f,r_i)\,+\,\Delta_S
\frac{\int \,d r'' \,\phi(r'')\,\,G^H_{\omega}(r_f,
r'')\,\int \,d r' \,\phi(r')\,\,G^H_{\omega}(r',r_i)\,}
{1\,\,-\,\,\Delta_S \int
\,d r'\,\,\int \,d r'' \,\phi(r')\,\,G^H_{\omega}(r', 
r'')\,\,\phi(r'')}\,.
\ee

The Green's function of \eq{MASTER1} has a quite different asymptotic 
behavior than $g\Le \omega,\,r_f,\,r_i\Ra$ due to the contribution of the denominator.
The zero of denominator generates a Regge pole ( resulting Pomeron ) in $\,\omega\,$. 
Therefore, the asymptotic behavior of Green's function of the  \eq{MASTER1}
depends on what is dominant in given kinematic region,
either $\,G^{H}_{\omega}\,$ or this pole.

First, we consider the region of rapidity, given by \eq{SPOM4}.
Now, we have two conditions on  $\Delta_{S-H}$, which should be
satisfied,
so that the  soft-hard pole contribution in \eq{MASTER1} will
dominate. 
First is the condition of
the existence  of soft pole solution in denominator of \eq{MASTER1}. This
condition is
given by  \eq{USL1}- \eq{INQ1}. Second , we need  to  check that this
soft-hard  contribution to the Green's function will give
a larger contribution than either the contribution of  
$\,G^{H}_{\omega}(r_f,r_i)\,$ or of the  numerator of the second term
of \eq{MASTER1}. From the derivation of the \eq{EQU3}-\eq{EQU6} is clear,
that the asymptotic behavior of both  functions  in the leading order  is
given by $e^{\tilde{\omega}\,Y}$. 
So, all these conditions  together
define the following restriction for the kinematic region for
resulting Pomeron :

\be\label{EQU11}
\Delta_{S-H}\,\geq\,\tilde{\omega}\,.
\ee

We have obtained this restriction before, see \eq{USL1}-\eq{USL2},
and can say again,  that this restriction is satisfied for the case of
very small $\bar{\alpha}_S\,$, when $\bar{\alpha}_S\,<\,0.1\,$
and therefore $\Delta_{S}\,>\,\tilde{\omega}_{S}\,\sim\,\tilde{\omega}\,$,
or for the case of large $\frac{r_i\,+\,r_f}{2}$, when
$\frac{\,r_f\,+\,r_i\,}{2}\,>\,
\frac{\omega_L\,r_H}{\Delta_{S-H}\,}$.
So, in both these cases asymptotic behavior of Green's function
will be defined by resulting pole.
Otherwise,  
we have  the answer of \eq{EQU6}:

\be\label{EQU66}
G(Y;r_f\,,r_i\,)\,=\,\
\sqrt{\,\frac{3\,r_i}{8\,\pi \,Y\,\omega_L\,r_H\,D}}\,\,
e^{ \tilde{\omega}\,Y\,\,-\,\,\frac{2}{9\,\sqrt{3}}
\frac{D\,( \,\tilde{\omega}\,Y\,)^3}{(r_f\,+\,r_i)^2\,/4}}\,
\ee
in the region of  rapidity:
\be \label{SPOM44}
0 \,<\,Y\,<\, \frac{r^2_f}{\omega_L\,r_H}\,\,.
\ee

 Another possibility for the resulting  Pomeron dominance is to have large
values of rapidity.
In this case the \eq{EQU66} is not valid anymore, and we have to use 
equations \eq{SPOM10}-\eq{SPOM12}.
So, using Eq.~\ref{SPOM11}, we can claim that for: 
\be
\Delta_{S-H}\,\ge\,\omega_{SP}
\ee
we obtain 
the constraint on the value  of rapidity for which the resulting Pomeron 
governs the high energy asymptotic:
\be
Y\,\ge\,\frac{\,\omega_L\,r_H}{\,D\,\Delta^{2}_{S}}\,
\ee

Let's assume now, that we have resulting Pomeron regime,
i.e. consider only the second term of \eq{MASTER1}.
We want to   rewrite our Green's function and estimate
the resulting Pomeron contribution.
For that,
first of all, we have to
expand the denominator of resulting Pomeron, Eq.~\ref{MASTER1}, 
with respect to $\Le\,\omega-\Delta_{S-H}\,\Ra $ : 

$$
1\,\,-\,\,\Delta_S \int
\,d r'\,\,\int \,d r'' \,\phi(r')\,\,G^H_{\omega}(r', 
r'')\,\,\phi(r'')\,\approx\,
1\,\,-\,\,\,\frac{\Delta_{S}}{\,\omega\,-\,\tilde{\omega}_{S}\,}\,\,\approx\,
$$
$$
1\,\,-\,\,\Le\,
\,\frac{\Delta_{S}}{\,\omega\,-\,\tilde{\omega}_{S}\,}\,
\Ra_{\omega=\Delta_{S-H}}\,
-\,\Delta_S\,\Le\,\omega\,-\,\Delta_{S-H}\,\Ra\,
\frac{\partial}{\partial\,\omega}
\Le\,
\,\frac{1}{\,\omega\,-\,\tilde{\omega}_{S}\,}\,\Ra_{\omega=\Delta_{S-H}}\,
+\,...\,\approx\,
\frac{\,\Le\,\omega\,-\,\Delta_{S-H}\,\Ra\,}{\Delta_{S}}\,\,.
$$

Here we take  $\delta\omega$ and  $\Delta_{S-H}$ from  \eq{EXP2}.
Using this expansion, we rewrite our 
Green's  function:

\be\label{SOFTIN} 
G^{DIS}(\,Y\,,r_f\,,r_i\,)\,\,=\,
\frac{\Delta_S}{\,q_f\,q_i\,}\,
\int^{a+i\,\infty}_{a-i\,\infty}\,\frac{d\,\omega}{2\,\pi\,i}\,
\Le\,\frac{s}{q_f\,q_i}\,\Ra^{\,\omega\,}\,
\frac{\,\tilde{g}\Le \omega,\,r_f,\,r_i\Ra\,}
{\omega\,-\,\Delta_{S-H}}\,\,,
\ee

where
$$
\tilde{g}\Le \omega,\,r_f,\,r_i\Ra\,=\,
\int \,d r'' \,\phi(r'')\,\,G^H_{\omega}(r_f,
r'')\,\int \,d r' \,\phi(r')\,\,G^H_{\omega}(r',r_i)\,
$$

Taking the contour integration we obtain the answer:

\be \label{SOFTAN}
G^{DIS}(\,Y\,,r_f\,,r_i\,)\,=\,
\frac{\Delta_S}{\,q_f\,q_i\,}\,
\Le\,\frac{s}{q_f\,q_i}\,\Ra^{\,\Delta_{S-H}\,}\,
\tilde{g}\Le \omega=\Delta_{S-H},\,r_f,\,r_i\Ra\,,
\ee

or using the 
Green's function of \eq{SP223} we have:

\be\label{S_H}
G^{DIS}\Le\,Y\,;r_i\,,\,r_f\,\Ra\,\approx\,
\frac{\Delta_S\,r_S^2}{\,q_f\,q_i\,}
\frac{\Delta_{S-H}^{-2/3}}{\Le\omega_L\,r_H\,D\,\Ra^{4/3}}
\Le\,\frac{s}{q_f\,q_i}\,\Ra^{\,\Delta_{S-H}\,}\,.
\ee

  So, from  this consideration, we see, 
that we  have the resulting, phenomenological,
Pomeron contribution, determined by \eq{S_H}, in the following cases:

\begin{itemize}

\item first, in the range of rapidity

$$
0 \,<\,Y\,<\, \frac{r^2_f}{\omega_L\,r_H}\,\,,
$$

for the small values of $\bar{\alpha}_S\,$: $\bar{\alpha}_S\,<\,0.1\,$;

\item second, in the range of rapidity

$$
0 \,<\,Y\,<\, \frac{r^2_f}{\omega_L\,r_H}\,\,,
$$  

for the large values of $\frac{r_i\,+\,r_f}{2}$:

$$
\frac{\,r_f\,+\,r_i\,}{\,2\,}\,>\,
\frac{\omega_L\,r_H}{\Delta_{S-H}\,}\,\sim\,r_S\,\,;
$$

\item third, for large rapidities:

$$
Y\,\ge\,\frac{\,\omega_L\,r_H}{\,D\,\Delta^{2}_{S}}\,.
$$
\end{itemize}

One can see from \eq{SOFTIN}  that the soft-hard transition generates a 
Pomeron satisfying the initial condition for the moments of 
the deep inelastic structure function.
Therefore, in the kinematic region, 
where the $\Delta_{S-H}$ contributes , the DGLAP 
evolution equations only affect the $Q^2$ dependence of the deep inelastic 
structure function, while the energy dependence is entirely determined by 
the soft Pomeron. Such a situation has been studied in details in Ref. 
\cite{LOP}.  In other kinematic regions  we have a normal
DGLAP evolution equation 
with the initial conditions which do not contain the soft-hard Pomeron 
contributions.

It should also be stressed that at fixed virtuality $Q^2$ at high energy, 
asymptotic behavior is 
determined by the exchange of the soft Pomeron independent 
of the  value of $Q^2$.  However, the coupling of the soft-hard Pomeron to the
photon depends on the value of the photon virtuality. 

 Estimating  our function, \eq{S_H}, for the case 
of large $Q^2$, 
we see that the contribution of the rightmost singularity 
( soft-hard Pomeron) 
is small at large $q_f\,$. 
This fact makes our life more complicated and we 
have to remember the infinite number of the poles  that we have in 
\eq{MASTER}. 
Indeed, these poles accumulate to zero, 
as  will be seen  latter,
and, therefore, we  face  the problem of an  essential singularity
at zero, which possibly may give a large contribution 
to the Green's function.
We have to estimate the contribution of 
these secondary poles in $G^{DIS}\Le\,Y\,;r_i\,,\,r_f\,\Ra $.
This estimate is considered in Appendix D, and the answer is

\be\label{SEC6}
G^{DIS}\,(\,Y\,;r_f\,,r_i\,)\,\,=\,\,
\frac{r_i\,2^{-1/3}}{\Delta_S\,\sqrt{\pi}\,D^{5/6}}\,
\sum_{n}\,(-1)^{n}\,
\frac{
((\tilde{Ai}(\xi;\zeta))_{r_i=r_S}\,
(\tilde{Ai}(\xi;\zeta))_{r_f=r_S})_{\omega=\omega_{n}}}{(\pi\,n)^{4/3}\,}\,
\,\,e^{\frac{2}{3}\,\left(\frac{4}{D}\right)^{\frac{1}{2}}\,
\frac{r_H\,\omega_L}{\pi\,n}\,
\ln\,\frac{s}{q_i\,q_f}}\,\,.
\ee

The  estimate of the value of this sum we may perform using the method of steepest descent. 
This estimate gives 
the $e^{\sqrt{\ln\,(s)}}$ behavior for the sum, and this contribution 
is smaller, than the contribution
from the right-most singularity, $\Delta_{S-H}$. Therefore, in our calculation,
we can only need consider the  resulting Regge pole
contribution and  neglect the contribution of the secondary poles.

\section{Conclusion}

This paper is a realization of old ideas to interface non-perturbative 
Pomeron and perturbative BFKL Pomeron \cite{LETA}. The new ingredient in
our approach is  that the typical scale for the ``soft" Pomeron is
considered \cite{KL,KKL}  to be rather large
($\,M_0\,\approx\,2\,\,GeV\,\,$ or
$r_S \,=\,
\ln(M^2_0/\Lambda^2)\,\approx\,\,4.6\,)\,$, larger than
the values of transverse momenta ($p_t$ for which we can use perturbative
BFKL approach ($p_t\,\geq\,1
\,GeV$,\,\,$ r_H\,\approx\,\ln(1GeV^2/\Lambda^2)\,=\,3.2$). In other words ,
we assume that $r_S\,\geq\,r_H$.  

We hoped that taking into account the running QCD coupling constant we would 
obtain the resulting singularity which would  govern ``soft" processes 
close to unity, as it appears in  the experimental observations \cite{DL}.
This hope is based on the fact that the BFKL Pomeron with running QCD
coupling leads  to a  singularity at a  angular momentum which is equal to
unity. However, it turns out that even for running $\as$ the resulting
singularity is  located to the right of $\omega_S\,
=\,\omega^{BFKL}\frac{r_H}{r_S}$. Using the next-to-leading order BFKL
estimates for $\omega^{BFKL} \,\approx\,0.2$ \cite{LF,CCNO,GPS}, we obtain
$\omega_S\,=\,0.139 $ which is  larger than the Pomeron intercept of the D-L
 phenomenological Pomeron  \cite{DL}.  The estimate of
the Pomeron intercept from the single inclusive production suggests a higher
value of the Pomeron intercept \cite{INCL} which is in better agreement
with our prediction.

The main result of this paper is concentrated in  \eq{EXP2} and \eq{USL44} which
 give
the position of the resulting singularity  for different ranges of $\as$ and
non-perturbative input $\Delta_S$.
 We also  discussed  the interplay between the BFKL
Pomeron
and the resulting Pomeron in deep inelastic processes which provide a possibility to
measure the hard BFKL Pomeron at large virtualities of photon, and to investigate the
interface between the  ``soft" and  the ``hard" Pomerons for small values of the photon
virtualities. To our  surprise our result shows (see \eq{S_H} ) that even  at
large
photon virtualities  the contribution of the resulting singularity dominantes at high
energy (low $x$ ).  
In other words we cannot consider DIS at low $x$ as a typical hard
process where we can safely use perturbative QCD. The admixture of non-perturbative
interaction is so essential that it changes the effective Pomeron intercept.

 It is  interesting  to compare our model with the model
introduced in Ref.\cite{CTM}, where  two Pomeron model
was also introduced.
We have no ideological disagreements  with this model but  we differ from it 
in the way a transition from soft to hard Pomerons occurrs  and in the  origin of
soft Pomeron.  We assumed that the existence of the soft Pomeron is related to typical
non-perturbative phenomena e.g. QCD instantons and/or the scale anomaly in QCD, while in
Ref.\cite{CTM}  M.~Ciafaloni, M.~Taiuti and A.~H.~Mueller
study a possibility that the non-perturbative Pomeron would be generated by the
behavior of the QCD coupling at low values of transverse momenta.
Our ``soft" Pomeron has normal Gribov diffusion in impact parameter space which we 
admixture with
diffusion in transverse momenta typical for the BFKL Pomeron.  The freezing of
the QCD coupling leads to a new picture which was suggested in Ref.\cite{CCSS1} and
which is certainly an alternative approach to ours.

 We would like to stress that in this paper we consider only the interface of two Pomerons:
the soft Pomeron and the BFKL Pomeron. The reality could be much richer and 
could possibly lead  to another 
scenario of interface the long distance interaction with the short distance QCD physics.
 As was 
pointed out in Ref. \cite{KKLS}, the soft Pomeron could be suppressed if it is built from QCD 
instantons, \cite{KKL}, in the high parton density environment. In other words, 
if the interface 
between long and short distances interaction passes the stage of high parton density QCD, 
the soft 
Pomeron does not contribute to the high energy asymptotic behavior of 
the scattering amplitude.

The main result of this paper
we consider to be  the observation that the non-perturbative correction could be essential for
so called ``hard" process  at
low $x$ and large scale of hardness ( like DIS at low $x$). 

{\bf  Acknowledgments:}
  
 We wish to thank  Jochen Bartels, Errol Gotsman and Uri Maor   for very
fruitful discussions on the subject.  We are very grateful to our referee for all his 
remarks which, we hope, improves our presentation.

 We would like to thank the DESY Theory
Group and the BNL Nuclear Theory Group  for their hospitality and creative atmosphere
during several
stages of this work.

 E.L.  is indebted to the Alexander-von-Humboldt Foundation for the
award that gave him a possibility to work on high energy  physics during the
last year.

 This research was supported in part by the
GIF grant \# I-620-22.14/1999, and by the Israel Science Foundation,
founded by the Israeli Academy of Science and Humanities.

 Special thanks go to the Dima Kharzeev and Yuri V. Kovchegov who took part
in the first part of the research and whose discussions and support were very
helpful.

\appendix
\section{}

We consider the mellin transform  over $\omega$ in \eq{ORIG5}
with the contour of integration 
defined in  a  usual way:
to  the right of  all singularities of the integrand.
First consider the limit $Y\,\rightarrow\,0 $:
\be\label{INTEG1}
G^{1}(0;,r_f\,, r_i\,)\,\,=\,\frac{2\,r_i\,\,}{\,r_i\,+r_f\,}
\,\,\sqrt{\frac{1}{\tilde{\omega}\,D}}\,\,
\,\int^{\tilde{\omega}+i\infty}_{\tilde{\omega}-i\infty}\,\,
\frac{d\omega}{2\,\pi\,i}\,\,
\frac{e^{-\,\delta\,r
\sqrt{\frac{\omega\,-\,\tilde{\omega}}{D\,\tilde{\omega}}}}}
{\sqrt{\omega\,-\,\tilde{\omega}}}\,\,.
\ee
Closing the contour  of integration to the left over the cut from the
$-\infty $
to $\tilde{\omega}$ with the variables change:
\be
\omega\,-\,\tilde{\omega}\,=\,e^{i\,\pi}\,\delta\omega
\ee 
on the upper edge  of the cut, and 
\be
\omega\,-\,\tilde{\omega}\,=\,e^{-i\,\pi}\,\delta\omega
\ee 
on the lower edge  of the cut. In so doing, the integral \eq{INTEG1}
turns into a sum:
\be
G^{1}(0;r_f\,, r_i\,)\,\,=\,-\,
\frac{2\,r_i\,}{\,r_i\,+r_f\,}\,\,
\sqrt{\frac{1}{\tilde{\omega}\,D}}\,
\int_{0}^{\infty}\,\,
\frac{d\delta\omega}{2\pi\,i}\,\,\,\frac{e^{-i\,\delta\,r
\sqrt{\frac{\delta\omega}{D\,\tilde{\omega}}}}}
{i\,\sqrt{\delta\omega\,}}\,\,-\,\,
\frac{2\,r_i\,}{\,r_i\,+r_f\,}\,\,
\sqrt{\frac{1}{\tilde{\omega}\,D}}\,
\,\int^{0}_{\infty}\,\,
\frac{d\delta\omega}{2\pi\,i}\,\,
\frac{e^{i\,\delta\,r
\sqrt{\frac{\delta\omega}{D\,\tilde{\omega}}}}}
{-i\,\sqrt{\delta\omega\,}}\,,
\ee
or
\be
G^{1}(0;r_f\,, r_i\,)\,\,=\,\frac{4\,r_i\,}{\,r_i\,+r_f\,}\,\int_{0}^{\infty}\,\,
\frac{d\delta\omega}{2\pi\,}\,\,
e^{-i\,\delta r\,\delta\omega}\,\,-\,\,\frac{4\,r_i\,}{\,r_i\,+r_f\,}\,
\,\,\int^{0}_{\infty}\,\,
\frac{d\delta\omega}{2\pi\,}\,\,
e^{i\,\delta r\,\delta\omega}\,\,,
\ee
which leads to (after  changing variable 
$\delta\omega\,\rightarrow\,-\,\delta\omega $ in the second term
of this expression)
\be\label{Delta}
G^{1}(0;r_f\,, r_i)\,\,=\,\frac{4\,r_i\,}{\,r_i\,+r_f\,}\,\int_{-\infty}^{\infty}\,\,
\frac{d\delta\omega}{2\pi\,}\,\,
e^{-i\,\delta r\,\delta\omega}\,\,=\,\,
\,\frac{4\,r_i\,}{r_i\,+r_f\,}\,\delta(\delta\,r)\,.
\ee

Coming  back to \eq{INTEG1} with non zero Y, we have
\be\label{INTEG11}
G^{1}(Y;r_f\,, r_i\,)\,\,=\,\frac{2\,r_i\,\,}{\,r_i\,+r_f\,}
\,\,\sqrt{\frac{1}{\tilde{\omega}\,D}}\,\,
\,\int^{\tilde{\omega}+i\infty}_{\tilde{\omega}-i\infty}\,\,
\frac{d\omega}{2\,\pi\,i}\,\,
\frac{e^{\omega\,Y\,-\,\delta\,r
\sqrt{\frac{\omega\,-\,\tilde{\omega}}{D\,\tilde{\omega}}}}}
{\sqrt{\omega\,-\,\tilde{\omega}}}\,\,.
\ee
This integral can be evaluated 
using the method of steepest decent, with the  the saddle point
in  $\omega$ determined by:
\be \label{SADDLEOM}
Y\,-\,\frac{\delta r}{2}\,\frac{1}{\sqrt{D\,\tilde{\omega}}}\,
\frac{1}{\sqrt{\omega\,-\,\tilde{\omega}}}\,=\,0\,,
\ee
In the limit $\,\,\omega\,>>\tilde{\omega}\,\,$,  \eq{SADDLEOM} leads to the
following saddle point value:
\be
\omega_{SP}\,=\,\,\Le\frac{\delta r}{2}\Ra^{2}\,
\frac{1}{Y^2}\,\frac{1}{\tilde{\omega}\,D}\,,
\ee
and to the asymptotic behavior of   the Green's function:
\be\label{NAN_1}
G^{1}(Y;r_f\,, r_i\,)\,=\,\sqrt{\frac{1}{2\,\pi\,Y\,\tilde{\omega}\,D}}\,
e^{\,-\,\frac{(\delta r)^2}{4\,Y\,\tilde{\omega}\,D}}\,\,,
\ee
which is valid for $\omega_{SP}\,>>\tilde{\omega}$, or 
$\omega\,\sim\,1/Y\,>>\,\tilde{\omega}$, or $Y\,<<\frac{1}{\tilde{\omega}}$.

Now we consider \eq{EXAMP}.
Here there are complex conjugates
functions and, therefore, the answer is the 
sum of the real parts of these functions:
$$
g(\omega,r_f, r_f)\,\,\propto\,
2\,\cos(\pi/6)\,\int_{0}^{\infty}\,\frac{d\,\nu}{\sqrt{\nu}}\,\,
e^{-\,\xi\,\nu\,\cos(\pi/3)\,\,-\,\,\frac{\nu^3}{3}\,}\,
\cos(\,\xi\,\nu\,\sin(\frac{\pi}{3}))\,+\,
$$
\be\label{EQU10}
2\,\sin(\pi/6)\,\int_{0}^{\infty}\,\frac{d\,\nu}{\sqrt{\nu}}\,\,
e^{-\,\xi\,\nu\,\cos(\pi/3)\,\,-\,\,\frac{\nu^3}{3}\,}\,
\sin(\,\xi\,\nu\,\sin(\frac{\pi}{3}))\,.
\ee
Because $\xi$ and $\delta\omega$ are  small,  
while  $\nu$ is the order of 1, we can expand
our function with respect to  $\xi\,\nu$- term.
Integrating this expansion over $\nu$
we  obtain  the following result for the first two orders of $\xi\,$:
\be\label{47}
g(\omega,r_f, r_f)\,\,\propto\,
\frac{\Gamma(\frac{1}{6})}{3^{2/3}}\,-\,
\xi^2\,\,
\frac{\Gamma(\frac{5}{6})}{8\,\,\,3^{4/3}}\,.
\ee
Taking into account expression for $\xi$:
$$
\xi\,=\,\Le\frac{4\omega}{D\omega_L\,\,r_H}\Ra^{1/3}\,
\frac{\frac{r_i\,+\,r_f}{2}}{\,\omega}\,(\omega\,-\,\tilde{\omega})\,\,,
$$
we obtain finally:
\be\label{EQU2_2}
g(\omega,r_f, r_f)\,\,\propto\,
\frac{\Gamma(\frac{1}{6})}{3^{2/3}}\,-\,
\Le\frac{4\omega}{D\omega_L\,\,r_H}\Ra^{2/3}\,
\frac{(\frac{r_i\,+\,r_f}{2})^2}{8\,\omega^2}\,(\delta\omega)^{2}\,
\frac{\Gamma(\frac{5}{6})}{3^{4/3}}\,.
\ee

This answer is also valid for the case when
$\delta\omega\,/\tilde{\omega}\,\sim\,1\,$ and $\xi\sim\,1$.
In this case, the series of \eq{47} will be convergent 
due to the $\frac{1}{\,n\,!}$ in $e^{-\xi\nu}$ expansion.
Indeed, we see from \eq{47}, taking $\xi\,=\,1$, that
$$
\frac{\Gamma(\frac{5}{6})}{\Gamma(\frac{1}{6})\,8\,\,\,3^{2/3}}\,
\propto\,\frac{1}{80}\,<<1\,\,,
$$
is small.
The following condition  for $\xi\,\,$  arises in this case:
$\,\,\xi\,\leq\,3$, and  we obtain the simple estimate
for the region of $\,\omega\,$ where the series is still convergent 
and the given solution is still valid :
\be\label{conver}
\frac{\omega}{\tilde{\omega}}\,\leq\,3\,\,,
\ee
where we still have   $\frac{\delta\omega\,}{\tilde{\omega}}\,\leq\,2\,\,.$
Now, if we go back to the \eq{ORIG5}, and taking 
$\omega\,=\,\tilde{\omega}$, we  see, that the correcting term
to \eq{ORIG5} is still small, 
$$
e^{\frac{D\,\tilde{\omega}}{\omega\,(\frac{r'\,+\,r''}{2})^{2}}}
\sim\,1\,+\,\frac{1}{96}\,<<1
$$
Therefore, in the case
$\delta\omega\,/\tilde{\omega}\,\sim\,1\,$ and $\xi\sim\,1$
we obtained overlapping solutions for Green's function, \eq{ORIG5} and \eq{47},
which match each other.
So, 
we now know the form of our Green's function in the whole region of positive $\xi$.

  We also find here the  $G(Y;r_f\,,r_i\,)$ function. 
Performing Mellin transform of \eq{EQU10} 
and integrating over $\delta\omega$ in the first approximation
in  $\xi\nu$ we have :
\be\label{EQU3}
G(Y;r_f\,,r_i\,)\,\propto\,
\int_{i\infty}^{-i\infty}\frac{d(\delta\omega)}{2\,\pi\,i}\,
e^{\tilde{\omega}\,Y\,+Y\,\delta\omega\,-
\nu\,\Le\frac{4\,}{D\omega_L\,\,r_H}\Ra^{1/3}\,\frac{r_i\,+\,r_f}{2\,\tilde{\omega}^{2/3}}\,
\,cos(\pi/3)\,\delta\omega\,}\,\int_{0}^{\infty}\,\frac{d\,\nu}{\sqrt{\nu}}\,
e^{-\frac{\nu^3}{3}}\,\,,
\ee
or
\be\label{EQU4}
G(Y;r_f\,,r_i\,)\,\propto\,
e^{\tilde{\omega}\,Y\,}
\int_{-\infty}^{\infty}\frac{d(\delta\omega)}{2\,\pi\,}\,
e^{i\,\delta\omega\,(Y\,-\nu\,
\Le\frac{4\,}{D\omega_L\,\,r_H}\Ra^{1/3}\,\frac{r_i\,+\,r_f}{2\,\tilde{\omega}^{2/3}}\,
\,cos(\pi/3)\,)}\,\int_{0}^{\infty}\,\frac{d\,\nu}{\sqrt{\nu}}\,
e^{-\frac{\nu^3}{3}}\,\,,
\ee
or
\be\label{EQU5}
G(Y;r_f\,,r_i\,)\,\propto\,
e^{\tilde{\omega}\,Y\,}
\int_{0}^{\infty}\,\frac{d\,\nu}{\sqrt{\nu}}\,
\delta\,(Y\,-\nu\,\Le\frac{4\,}{D\omega_L\,\,r_H}\Ra^{1/3}\,
\frac{r_i\,+\,r_f}{2\,\tilde{\omega}^{2/3}}\,
\,cos(\pi/3)\,)\,e^{-\frac{\nu^3}{3}}\,.
\ee
The final answer, obtained for this Green's function in 
this asymptotic limit, is:
\be\label{EQU6_1}
G(Y;r_f\,,r_i\,)\,=\,\
\sqrt{\,\frac{3\,r_i}{8\,\pi \,Y\,\omega_L\,r_H\,D}}\,\,
e^{ \tilde{\omega}\,Y\,\,-\,\,\frac{2}{9\,\sqrt{3}}
\frac{D\,( \,\tilde{\omega}\,Y\,)^3}{(r_f\,+\,r_i)^2\,/4} }\,.
\ee

\section{}
 The \eq{SHT01} has the following form :

\be\label{Ap1}
G^{S-H}(r_f,r_i)\,=\,D(\omega;r_f,r_i)+
\int\int\,D(\omega;r_f,r_a)\Le\,K^{S}(r_a,r_b)+K^{H}(r_a,r_b)\,\Ra\,
G^{S-H}(r_b,r_i)\,d^2 r_a\,d^2 r_b\,.
\ee

We are searching  the solution of \eq{Ap1} in the form

\be\label{Ap2}
G^{S-H}(r_f,r_i)\,=\,\int\,G^{H}(r_f,r_a)\,f(r_a,r_i)\,d^2\,r_a\,\,.
\ee

Inserting \eq{Ap2} in \eq{Ap1} we obtain:

\be\label{Ap3}
\int\,G^{H}(r_f,r_a)\,f(r_a,r_i)\,d\,r_a\,=\,
\frac{\delta^2(r_f\,-\,r_i)}{\omega}\,+\,
\frac{\Delta_S}{\omega}\,\phi(r_f)\int\,\int\,
\phi(r_a)G^{H}(r_a,r_b)f(r_b,r_i)\,d^2\,r_a\,d^2\,r_b\,+\,
\ee
$$
\,+\,
\int\,\int\,\int\,D(\omega;r_f,r_a)\,K^{H}(r_a,r_b)\,
G^{S-H}(r_b,r_c)\,f(r_c,r_i)\,d^2 r_a\,d^2 r_b\,d^2 r_c\,\,,
$$

where we used that
$K^{S}(r_f,r_i)\,=\,\Delta_S\,\phi(r_f)\,\phi(r_i)$ and
$D(\omega;r_f,r_i)\,=\,\frac{\delta^2(r_f\,-\,r_i)}{\omega}\,$.

And because (see \eq{GBFKL})

\be\label{Ap4}
G^{H}(r_f,r_i)\,=\,D(\omega;r_f,r_i)\,+\,
\int\,\int\,D(\omega;r_f,r_a)\,K^{H}(r_a,r_b)\,
G^{H}(r_b,r_i)\,\,d^2 r_a\,d^2 r_b\,\,
\ee

we have:

\be\label{Ap5}
\int\,G^{H}(r_f,r_a)\,f(r_a,r_i)\,d\,r_a\,=\,
\frac{\delta^2(r_f\,-\,r_i)}{\omega}\,+\,
\frac{\Delta_S}{\omega}\,\phi(r_f)\int\,\int\,
\phi(r_a)G^{H}(r_a,r_b)f(r_b,r_i)\,d^2\,r_a\,d^2\,r_b\,+\,
\ee
$$
\,+\,
\int\,\Le\,G^{H}(r_f,r_a)\,-\,\frac{\delta^2(r_f\,-\,r_a)}{\omega}\,\Ra\,
f(r_a,r_i)\,d^2 r_a\,\,.
$$
 
This equation leads to

\be\label{Ap6}
f(r_f,r_i)\,=\,\delta^2(r_f\,-\,r_i)\,+\,
\Delta_S\,\phi(r_f)\int\,\int\,
\phi(r_a)G^{H}(r_a,r_b)f(r_b,r_i)\,d^2\,r_a\,d^2\,r_b\,.
\ee

Multiplying both parts of \eq{Ap5} on the integral
$\,\int\,G^{H}(r_f,r_a)\phi(r_a)\,d^2\,r_a\,$ and integrating
over $\,r_f\,$ we obtain:

\be\label{Ap7}
\int\,\int\,
\phi(r_a)G^{H}(r_a,r_b)f(r_b,r_i)\,d^2\,r_a\,d^2\,r_b\,=\,
\int\,G^{H}(r_a,r_i)\phi(r_a)\,d^2\,r_a\,+\,
\ee
$$
\,+\,\Delta_S\,
\int\,\int\,\phi(r_a)\,G^{H}(r_a,r_b)\,\phi(r_b)\,d^2\,r_a\,d^2\,r_b\,
\int\,\int\,\phi(r_a)\,G^{H}(r_a,r_b)\,f(r_b,r_i)\,d^2\,r_a\,d^2\,r_b\,\,.
$$

\eq{Ap7} gives:

\be\label{Ap8}
\int\,\int\,
\phi(r_a)G^{H}(r_a,r_b)f(r_b,r_i)\,d^2\,r_a\,d^2\,r_b\,=\,
\frac{\int\,G^{H}(r_a,r_i)\phi(r_a)\,d^2\,r_a\,}
{1\,-\,\Delta_S\,
\int\,\int\,\phi(r_a)\,G^{H}(r_a,r_b)\,\phi(r_b)\,d^2\,r_a\,d^2\,r_b\,}.
\ee

Now, putting back \eq{Ap8} into \eq{Ap6} we obtain:

\be\label{Ap9}
f(r_f,r_i)\,=\,\delta^2(r_f\,-\,r_i)\,+\,
\frac{\Delta_S}{\omega}\,
\frac{\phi(r_f)\,\int\,G^{H}(r_a,r_i)\phi(r_a)\,d^2\,r_a\,}{
1\,-\,\Delta_S\,
\int\,\int\,\phi(r_a)\,G^{H}(r_a,r_b)\,\phi(r_b)\,d^2\,r_a\,d^2\,r_b\,}.
\ee

And, finally, we obtain for the solution of \eq{Ap1}

\be\label{Ap10}
G^{S-H}(r_f,r_i)\,=\,G^{H}(r_f,r_i)\,+\,
\Delta_S\,
\frac{\int\,G^{H}(r_f,r_a)\phi(r_a)\,d^2\,r_a\,
\int\,G^{H}(r_a,r_i)\phi(r_a)\,d^2\,r_a\,}
{1\,-\,\Delta_S\,
\int\,\int\,\phi(r_a)\,G^{H}(r_a,r_b)\,\phi(r_b)\,d^2\,r_a\,d^2\,r_b\,}\,.
\ee

\section{}

We search the poles which are the zeroes of denominator

\be \label{App1}
1\,\,-\,\,\Delta_S
\int \,d r'\,\,\int \,d r'' \,\phi(r')\,\,G^H_{\omega}(r',
r'')\,\,\phi(r'')\,\,\,=\,\,\,0\,\,.
\ee

To calculate  $G^H_{\omega}( r',r'')$ 
in this equation, we  use the result of \eq{ORIG5} and \eq{EQU2} for
different regions of $\omega$. For $\omega\,>>\,\tilde{\omega}$ we can use
the following asymptotic expression:
\be\label{NAN1}
G^{1}_{\omega}(r', r'')\,\,=\,\,
\frac{\,r''\,}{\,\tilde{r}^{1/2}}\,\,
\sqrt{\frac{1}{\omega\,\omega_L\,r_H\,D}}\,\,
e^{-\,\delta\,r\,\sqrt{\frac{\omega\,-\,\tilde{\omega}}{D\,\tilde{\omega}}}
+\frac{1}{96}\,\frac{D\,\tilde{\omega}}{\omega\,(\frac{r'\,+\,r''}{2})^2}}\,\,,
\ee
while for $\omega \,\sim \,\tilde{\omega}$, $\xi\,<<\,1\,,$
$\,\frac{\delta\omega\,}{\omega}\,<<\,\,1\,$, as well as for
$\,\xi\,\sim\,1\,,$ $\,\frac{\delta\omega\,}{\omega}\,\sim\,\,1\,$ we can
use the   asymptotic answer for Green's function  given by
\eq{EQU2}:
\be \label{SP223}
G^{2}_{\omega}(r', r'')\,\,=\,\,
\frac{\,r''\,}{4\,}\,\,
\sqrt{\frac{1}{\,\omega\,\omega_L\,r_H\,D}}\,\,
\Le\frac{4\,\omega}{\,\omega_L\,r_H\,D}\,\Ra^{1/6}\,
\Le
\frac{\Gamma(\frac{1}{6})}{3^{2/3}}\,-\,
\Le\frac{4\omega}{D\omega_L\,\,r_H}\Ra^{2/3}\,
\frac{(\frac{r'\,+\,r''}{2})^2}{8\,\omega^2}\,(\delta\omega)^{2}\,
\frac{\Gamma(\frac{5}{6})}{3^{4/3}}\,\Ra\,.
\ee

Let us now consider  \eq{App1} using  \eq{NAN1} for $G^H_{\omega}( r',r'')$:
\be\label{SHT21}
1\,\,-\,\,\Delta_S\,
\int \,d r'\,\,\int \,d r'' \,\phi(r')\,\,
\,\,G^{1}_{\omega}(r',r'')\,\,\,\phi(r'')\,\,\,=\,\,\,0\,\,.
\ee
Using new variables
\be
x\,=\,\frac{r'\,+\,r''}{2}\,,\,\,\,\,\,y=\frac{r'\,-\,r''}{2}\,\,,
\ee 
in  \eq{SHT21},
we  obtain: 
\be \label{SS}
\,\frac{\Delta_{S}}{\omega\,-\,\tilde{\omega}_{S}}\,
e^{\frac{1}{96}\,\frac{D\,\tilde{\omega}_S}{\omega\,r_{S}^2}}\,
\sqrt{\frac{\omega\,-\,\tilde{\omega}}{D\,\tilde{\omega}}}\,
\int_{0}^{\infty}\,dy\,\int_{-\infty}^{\infty}\,dx\,
e^{-y\,\sqrt{\frac{\omega\,-\,\tilde{\omega}}{D\,\tilde{\omega}}}}\,
e^{-\frac{2\Lambda^2}{q_{S}^{2}}\,\cosh\,(y)\,e^{x}}\,\,.
\ee
Due to the properties of the $\,\phi\,$ functions, the main contribution
in the integral comes from the region where $\,r\,\propto\,r_s\,$ and, therefore,
in \eq{SS}  we have used $\,\frac{r'+r''}{2}\approx\,r_s\,$ and
we defined there $\tilde{\omega}_S\,=\,\frac{\omega_{L}\,r_H}{r_S}$.
In the limit of small $\tilde{\omega}\,\rightarrow\,0$ we have 
$y\propto\tilde{\omega}^{1/2}\rightarrow\,0$ and, therefore, we can write
the integral in the form:
\be
\,\frac{\Delta_{S}}{\delta\omega}\,
e^{\frac{1}{96}\,\frac{D\,\tilde{\omega}_S}{\omega\,r_{S}^2}}\,
\sqrt{\frac{\omega\,}{D\,\tilde{\omega}}}\,
\int_{0}^{\infty}\,dy\,
e^{-y\,\sqrt{\frac{\omega\,}{D\,\tilde{\omega}}}}\,
\int_{-\infty}^{\infty}\,dx\,
e^{-\frac{2\Lambda^2}{q_{S}^{2}}\,e^{x}}\,\,,
\ee
or
\be\label{EXP1}
\,\frac{\Delta_{S}}{\,\delta\omega\,}\,
e^{\frac{1}{96}\,\frac{D\,\tilde{\omega}_S}{\omega\,r_{S}^2}}\,
\int_{0}^{\infty}\,dy\,e^{-y}\,
\int\,dx\,\phi(x)^{2}\,=\,\frac{\Delta_{S}}{\,\delta\omega\,}\,
e^{\frac{1}{96}\,\frac{D\,\tilde{\omega}_S}{\omega\,r_{S}^2}}\,.
\ee
The equation for the  intercept of the resulting  pole is :
\be
1\,-\,\,\frac{\Delta_{S}}{\,\omega\,-\,\tilde{\omega}_{S}}\,
e^{\frac{1}{96}\,\frac{D\,\tilde{\omega}_S}{\omega\,r_{S}^2}}\,=\,0\,,
\ee
or 
\be
1\,-\,\,\frac{\Delta_{S}}{\,\omega\,-\,\tilde{\omega}_{S}}\,
\Le\,1+\frac{D\,\tilde{\omega}_S}{96\,\omega\,r_{S}^2}\Ra\,=\,0\,.
\ee
This leads to the solution:
\be\label{App2}
\omega\,=\,\Delta_{S-H}\,=\,\Delta_{S}\,+\,\tilde{\omega}_{S}\,+\,
\frac{D\,\tilde{\omega}_S\,\Delta_S}{96\,r_{S}^2\,(\tilde{\omega}_S\,+\,
\Delta_S)}\,.
\ee

In  another region of $\omega$ we assume that
the position of the resulting pole $\omega \,=\,\Delta_{S-H}$ may be    
close to
$\tilde{\omega}_S\,$, $
\,\Delta_{S-H}\,=\,\tilde{\omega}_S\,+\,\delta\omega\,$,
where for  $\delta\omega$ there are two asymptotic limits. 
The first is the small $\delta\omega\,\,$, 
$\frac{\delta\omega\,}{\,\tilde{\omega}_S}\,<<\,\,1\,$, and 
the second for the $\delta\omega\,$ of the order of $\tilde{\omega}_S\,\,$,
$\frac{\delta\omega\,}{\,\tilde{\omega}_S}\,\sim\,\,1\,\,$.
We solve \eq{SHT2}by using
the Green function given by \eq{SP223} . Due to the form 
of functions $\phi (r)$ the main contribution in integral
of \eq{MASTER} comes from these $r'\,\approx\,r''\,\approx\,r_s$ .
Therefore, we  perform the integration in \eq{MASTER} only over the 
$\phi(r)$ functions, neglecting the term proportional to $\delta r $ in the
integrand. Using the properties of $\phi$ function, we obtain
\be
\int \,d r'\,\,\int \,d r'' \,\phi(r')\,\,\phi(r'')\,\,\,\approx\,\,4\,.
\ee
This follows from the fact that  
$$
\int \,d r\,\,\phi(r)^{2}\,\,=\,\,
A\,\int\,\frac{dq^2}{q_{S}^2}\,e^{-q^{2}/q_{S}^2}
=\,\,1\,.
$$
When  
$\delta\omega$ may be small or of the order $\tilde{\omega}_S\,$,  
( $\omega\,=\tilde{\omega}_S\,+\,\delta\omega\,$) , i.e.
$\frac{\delta\omega\,}{\,\tilde{\omega}_S}\,<<\,\,1\,$ or
$\frac{\delta\omega\,}{\,\tilde{\omega}_S}\,\sim\,\,1\,,$
we obtain the  following equation for resulting  pole:
\be \label{USL4}
1\,-\,\Delta_S\,r_S\,
\sqrt{\frac{1}{\,\omega\,\omega_L\,r_H\,D}}\,\,
\Le\frac{4\,\omega}{\,\omega_L\,r_H\,D}\,\Ra^{1/6}\,
\Le
\frac{\Gamma(\frac{1}{6})}{3^{2/3}}\,-\,
\Le\frac{4\omega}{D\omega_L\,\,r_H}\Ra^{2/3}\,
\frac{r_{S}^{2}}{8\,\omega^2}\,(\delta\omega)^{2}\,
\frac{\Gamma(\frac{5}{6})}{3^{4/3}}\,\Ra\,=\,0.
\ee
The solution of this equation  for the 
$\omega\,=\,\tilde{\omega}_S\,+\,\delta\omega\,$ with 
$\delta\omega \,\sim \,\tilde{\omega}_S\,\,$, up to the order 
$\Le\frac{\delta\omega}{\omega\,}\Ra^{2}\,,$ is :
\be\label{App3}
\omega\,=\Delta_{S-H}\,=\,\tilde{\omega}_S\,+\,\delta\omega\,=\,
\frac{2\,\Delta_{S}^{3}\,r_S\,}{\tilde{\omega}_{S}^{2}\,D^{2}\,}\,
\Le\,\frac{\Gamma(\frac{1}{6})}{3^{2/3}}\,\Ra^{3}\,\,.
\ee

\section{}

We again  consider our variable $\xi $:
$$
\xi\, = \, \left( \frac{4 \,\omega}{D \,\omega_L\,r_H}
\right)^{\frac{1}{3}}\,\{
\frac{r_f\,+\,r_i}{2}\,-\,\frac{\omega_L\,r_H}{\omega}\,\}\,.
$$
At  small values of $\omega$ we have:
\be
\xi\,\approx\, -\,\left( \frac{4 \,}{D\,}
\right)^{\frac{1}{3}}\,
\left(\frac{\omega_L\,r_H}{\omega}\right)^{\frac{2}{3}}\,.
\label{AM5A1}
\ee

We should  calculate the contribution of the secondary poles
to the second term of the Green function of  \eq{MASTER1}:

$$
G^{DIS}_{\omega}(\,r_f\,,r_i\,)\,\,=\,\,
\frac{\,\tilde{g}\Le \omega,\,r_f,\,r_i\Ra\,}
{1\,\,-\,\,\Delta_S \int
\,d r'\,\,\int \,d r'' \,\phi(r')\,\,G^H_{\omega}(r', 
r'')\,\,\phi(r'')}\,\,,
$$

We rewrite this equation in the following form:

\be\label{SEC1}
G^{DIS}_{\omega}(\,r_f\,,r_i\,)\,\,\approx\,\,
\frac{r_i,r_S}{4\,\pi\,\omega\,\omega_L\,r_H\,D}\,\left(
\frac{4\,\omega}{\omega_L\,r_H\,D}\right)^{\frac{1}{3}}\,
\frac{(\tilde{Ai}(\xi;\zeta))_{r_i=r_S}\,(\tilde{Ai}(\xi;\zeta))_{r_f=r_S}}
{1\,-\,\Delta_S\,\frac{\,r_S}
{\sqrt{\,4\,\pi\,r_H\,D\,\omega_L\,\omega}}\,\,\left(
 \frac{4 \,\omega}{D\,\omega_L\,r_H} \right)^{\frac{1}{6}}\,
\,\tilde{Ai}(\xi;\zeta=0)}
\ee

In the limit of small $\omega$ we can neglect 
1 in the denominator of \eq{SEC1}:

\be\label{SEC2}
G^{DIS}_{\omega}(\,r_f\,,r_i\,)\,\approx\,
-\,\frac{r_i\,\Delta_{S}^{-1}}{\sqrt{4\,\pi\,\omega\,\omega_L\,r_H\,D}}\,\left(
\frac{4\,\omega}{\omega_L\,r_H\,D}\right)^{\frac{1}{6}}\,
\frac{(\tilde{Ai}(\xi;\zeta))_{r_i=r_S}\,(\tilde{Ai}(\xi;\zeta))_{r_f=r_S}}
{\tilde{Ai}(\xi;\zeta=0)}\,.
\ee

The secondary pole positions are defined
by the zeroes of the $\tilde{Ai}(\xi;\zeta=0)$ function in the limit
$\omega\,\rightarrow\,0$.
For $\xi$ given by the \eq{AM5A1} we use the method of steepest descent
to estimate this function.
Calculated  saddle points are equal to  $\pm\,i\,\sqrt{\xi}$. They lead
to contribution: 

\be
\tilde{Ai}(\xi;\zeta=0)_{\omega\rightarrow\,0}\approx\,
\,\sqrt{\frac{\pi}{|\xi|}}\sin (\,\frac{2}{3}|\xi|^{3/2}\,)\,,
\ee

which together with \eq{SEC2} gives:

\be\label{SEC3}
G^{DIS}_{\omega}(\,r_f\,,r_i\,)\,\approx\,
-\frac{\,r_i\,\Delta_{S}^{-1}\,\pi^{-1/2}}
{(\,4\,\omega_L\,r_H\,D^2\,\omega^2\,)^{1/3}}\,
\frac{(\tilde{Ai}(\xi;\zeta))_{r_i=r_S}\,(\tilde{Ai}(\xi;\zeta))_{r_f=r_S}}
{\sin (\,\frac{2}{3}|\xi|^{3/2}\,)}\,.
\ee

We obtain for the secondary pole positions the following equation
( see also \cite{EL},\cite{haak}) :

\be
\sin (\,\frac{2}{3}\,\left(\, \frac{4 \,}{D\,}
\right)^{\frac{1}{2}}\,
\frac{\omega_L\,r_H}{\omega}\,)=0\,\,,
\ee

which leads to
\be\label{sep}\,
\omega_n\,=\,\frac{2}{3}\,\left(\, \frac{4 \,}{D\,}
\right)^{\frac{1}{2}}\,\frac{\omega_L\,r_H}{\pi\,n\,}.
\ee
 
Expanding $\,\sin\,$ in \eq{SEC3} in the vicinity of  $\omega=\omega_{n}$
for each $n$, we obtain:

\be\label{SEC4}
G^{DIS}_{\omega_{n}}(\,r_f\,,r_i\,)\,\approx\,
\frac{r_i\,2^{-1/3}}{\Delta_S\,\sqrt{\pi}\,D^{5/6}}\,
\frac{(\tilde{Ai}(\xi;\zeta))_{r_i=r_S}\,(\tilde{Ai}(\xi;\zeta))_{r_f=r_S}}
{(\pi\,n)^{4/3}\,(\omega\,-\omega_{n})}\,.
\ee

Integrating over $\omega$ and closing the  contour in $\omega$ over the
positions 
of the secondary poles
we obtain the answer for the contribution
of a  separate secondary pole:

\be\label{SEC5}
G^{DIS}_{n}(Y;\,r_f\,,r_i\,)\,\approx\,
\frac{r_i\,2^{-1/3}}{\Delta_S\,\sqrt{\pi}\,D^{5/6}}\,
\frac{(-1)^{n}\,
((\tilde{Ai}(\xi;\zeta))_{r_i=r_S}\,
(\tilde{Ai}(\xi;\zeta))_{r_f=r_S})_{\omega=\omega_{n}}}{(\pi\,n)^{4/3}\,}\,
\,e^{\frac{2}{3}\,\left(\frac{4}{D}\right)^{\frac{1}{2}}\,
\frac{r_H\,\omega_L}{\pi\,n}\,
\ln\,\frac{s}{q_i\,q_f}}\,\,.
\ee

The total contribution of the secondary poles is the sum over all 
$\,''\,n\,''\,$:

\be\label{SEC6_1}
G^{DIS}\,(\,Y\,;r_f\,,r_i\,)\,\,=\,\,
\frac{r_i\,2^{-1/3}}{\Delta_S\,\sqrt{\pi}\,D^{5/6}}\,
\sum_{n}\,(-1)^{n}\,
\frac{
((\tilde{Ai}(\xi;\zeta))_{r_i=r_S}\,
(\tilde{Ai}(\xi;\zeta))_{r_f=r_S})_{\omega=\omega_{n}}}{(\pi\,n)^{4/3}\,}\,
\,\,e^{\frac{2}{3}\,\left(\frac{4}{D}\right)^{\frac{1}{2}}\,
\frac{r_H\,\omega_L}{\pi\,n}\,
\ln\,\frac{s}{q_i\,q_f}}\,\,.
\ee

\end{document}